\documentclass[conference]{IEEEtran}
\usepackage{fancyhdr}
\IEEEoverridecommandlockouts
\usepackage{orcidlink}
\usepackage{cite}
\usepackage{amsmath,amssymb,amsfonts}
\usepackage{algorithm}
\usepackage{algpseudocode}
\usepackage{graphicx}
\usepackage{textcomp}
\usepackage{xcolor}
\usepackage{stfloats}
\usepackage{url}
\usepackage{verbatim}
\usepackage{xcolor}
\usepackage{ifthen}
\usepackage{multirow}
\usepackage{cleveref}
\usepackage{caption}
\usepackage{subcaption}
\usepackage{booktabs}
\usepackage[normalem]{ulem}
\usepackage{mathtools}
\usepackage{enumitem}
\usepackage{eso-pic}
\usepackage{pdfpages}

\def\BibTeX{{\rm B\kern-.05em{\sc i\kern-.025em b}\kern-.08em
    T\kern-.1667em\lower.7ex\hbox{E}\kern-.125emX}}

\newcommand{\showcomments}{yes}

\newcommand{\fixme}[1]{
    \ifthenelse{\equal{\showcomments}{yes}}{\textcolor{red}{[fixme: #1]}}{\ignorespaces}
}

\newcommand{\zz}[1]{
    \ifthenelse{\equal{\showcomments}{yes}}{\textcolor{blue}{[zz: #1]}}{\ignorespaces}
}

\newcommand{\jj}[1]{
    \ifthenelse{\equal{\showcomments}{yes}}{\textcolor{cyan}{[jj: #1]}}{\ignorespaces}
}

\newcommand{\niklas}[1]{
    \ifthenelse{\equal{\showcomments}{yes}}{\textcolor{purple}{[niklas: #1]}}{\ignorespaces}
}

\newcommand{\yixiao}[1]{
    \ifthenelse{\equal{\showcomments}{yes}}{\textcolor{orange}{[yixiao: #1]}}{\ignorespaces}
}

\newcommand{\simon}[1]{
    \ifthenelse{\equal{\showcomments}{yes}}{\textcolor{olive}{[simon: #1]}}{\ignorespaces}
}

\newcommand{\andrea}[1]{
    \ifthenelse{\equal{\showcomments}{yes}}{\textcolor{violet}{[andrea: #1]}}{\ignorespaces}
}

\newcommand{\erik}[1]{
    \ifthenelse{\equal{\showcomments}{yes}}{\textcolor{teal}{[erik: #1]}}{\ignorespaces}
}

\newcommand{\giulia}[1]{
    \ifthenelse{\equal{\showcomments}{yes}}{\textcolor{magenta}{[giulia: #1]}}{\ignorespaces}
}

\AddToShipoutPictureBG*{%
  \AtPageUpperLeft{%
    \put(\LenToUnit{\paperwidth-0cm},\LenToUnit{-1.2cm}){%
      \makebox[0pt][r]{%
        \includegraphics[height=1.2cm]{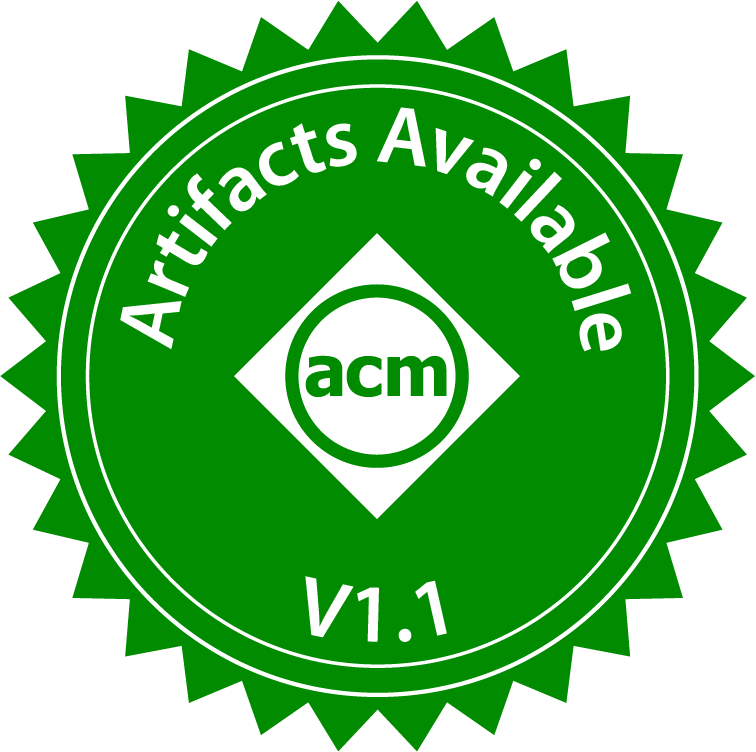}%
        \includegraphics[height=1.2cm]{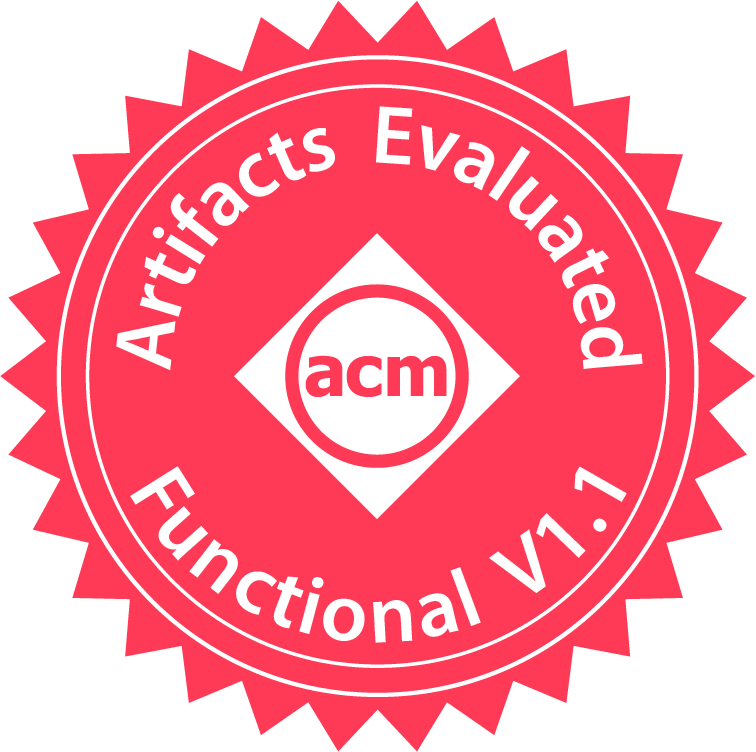}%
        \includegraphics[height=1.2cm]{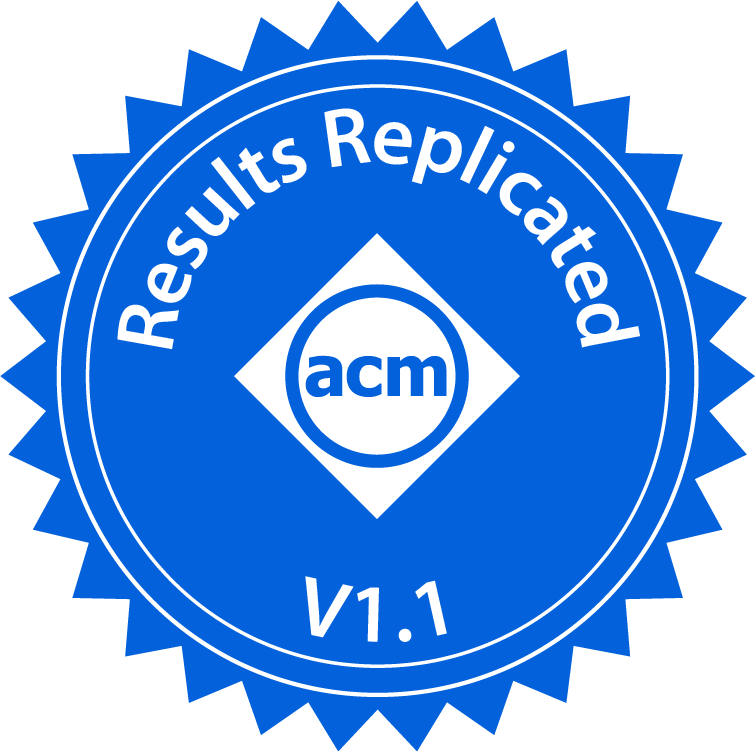}%
      }%
    }%
  }%
}

\begin{document}

\title{Rapid GPU-Based Pangenome Graph Layout}

\author{

\IEEEauthorblockN{Jiajie Li}
\IEEEauthorblockA{ 
    Cornell University\\
    Ithaca, NY, USA \\
    jl4257@cornell.edu \orcidlink{0000-0002-6775-2843}
    }
\and

\IEEEauthorblockN{Jan-Niklas Schmelzle}
\IEEEauthorblockA{
    Technical University of Munich \\
    Munich, Germany \\
    ge75qew@tum.de \orcidlink{0000-0001-8566-4049}
    }
\and

\IEEEauthorblockN{Yixiao Du}
\IEEEauthorblockA{
    Cornell University\\
    Ithaca, NY, USA \\
    yd383@cornell.edu \orcidlink{0000-0002-6106-1283}
    }
\and

\IEEEauthorblockN{Simon Heumos}
\IEEEauthorblockA{
    University of Tübingen\\
    Tübingen, Germany \\
    simon.heumos@qbic.uni-tuebingen.de \orcidlink{0000-0003-3326-817X}
    }
\and

\IEEEauthorblockN{Andrea Guarracino}
\IEEEauthorblockA{
    UTHSC \\
    Memphis, TN, USA \\
    aguarra1@uthsc.edu \orcidlink{0000-0001-9744-131X}
    }
\and

\IEEEauthorblockN{Giulia Guidi}
\IEEEauthorblockA{ 
    Cornell University\\ 
    Ithaca, NY, USA \\
    gg434@cornell.edu \orcidlink{0000-0001-8925-3239}
    }
\and

\IEEEauthorblockN{Pjotr Prins}
\IEEEauthorblockA{
    UTHSC \\
    Memphis, TN, USA \\
    jprins@uthsc.edu \orcidlink{0000-0002-8021-9162}
    }
\and

\IEEEauthorblockN{Erik Garrison}
\IEEEauthorblockA{
    UTHSC \\
    Memphis, TN, USA \\
    egarris5@uthsc.edu \orcidlink{0000-0003-3821-631X}
    }
\and

\IEEEauthorblockN{Zhiru Zhang}
\IEEEauthorblockA{ 
    Cornell University\\
    Ithaca, NY, USA \\
    zhiruz@cornell.edu \orcidlink{0000-0002-0778-0308}
    }

}

\maketitle

\thispagestyle{fancy}
\lhead{}
\rhead{}
\chead{}
\lfoot{\footnotesize{
SC24, November 17-22, 2024, Atlanta, Georgia, USA
\newline 979-8-3503-5291-7/24/\$31.00 \copyright 2024 IEEE}}
\rfoot{}
\cfoot{}
\renewcommand{\headrulewidth}{0pt}
\renewcommand{\footrulewidth}{0pt}


\begin{abstract}
Computational Pangenomics is an emerging field that studies genetic variation using a graph structure encompassing multiple genomes. Visualizing pangenome graphs is vital for understanding genome diversity. Yet, handling large graphs can be challenging due to the high computational demands of the graph layout process. 

In this work, we conduct a thorough performance characterization of a state-of-the-art pangenome graph layout algorithm, revealing significant data-level parallelism, which makes GPUs a promising option for compute acceleration. However, irregular data access and the algorithm's memory-bound nature present significant hurdles. To overcome these challenges, we develop a solution implementing three key optimizations: a cache-friendly data layout, coalesced random states, and warp merging. Additionally, we propose a quantitative metric for scalable evaluation of pangenome layout quality.

Evaluated on 24 human whole-chromosome pangenomes, our GPU-based solution achieves a 57.3x speedup over the state-of-the-art multithreaded CPU baseline without layout quality loss, reducing execution time from hours to minutes.

\end{abstract}


\section{Introduction}
\label{sec:introduction}

Low-cost genome sequencing~\cite{cheng2023scalable,rautiainen2023telomere} has made it possible to collect extensive genetic data for specific species, providing opportunities for deeper exploration.
Pangenomics~\cite{pangenome-graphs} is an emerging field of genomics that aims to understand the complete picture of the genetic variation of a species by studying multiple genomes~\cite{paten2017genome,golicz2020pangenomics}. Graphical pangenomics models a pangenome as a graph. 
This graph-based approach complements traditional reference-based genomics by revealing overlooked genetic variation when a single reference genome is used~\cite{ballouz2019time}.
In particular, the recent release of the first draft of the human pangenome reference~\cite{draft_human_pangenome_reference} represents a major milestone. 
This achievement represents a significant advance in human genetics, echoing the first release of the human genome sequence in 2001~\cite{first-human-genome}.

Pangenomes~\cite{wang2022human} can model the entire genomic variation of a given population~\cite{eisenstein2023every}. 
A variant refers to the differences between different genome sequences and can provide biological insights, such as disease susceptibility~\cite{groza2023geneticdiseases,yang2023neisseriameningitidis}, genome functionality identification~\cite{Guarracino2023}, and evolutionary studies~\cite{hubner2022we}. 

\begin{figure}[!htbp]
    \centering
    
    \begin{subfigure}[b]{0.53\linewidth}
        \centering
        \includegraphics[width=\linewidth]{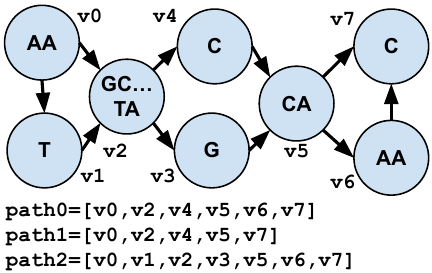}
        \caption{Variation graph data structure.}
        \label{fig:variation-graph-abstract}
    \end{subfigure}   
    \begin{subfigure}[b]{0.45\linewidth}
        \centering
        \includegraphics[width=\linewidth]{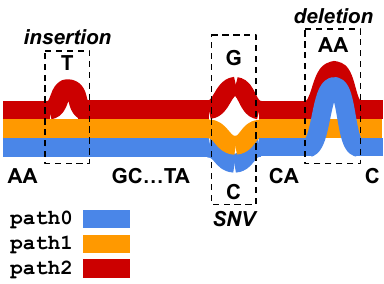}
        \caption{Visualization.}
        \label{fig:variation-graph-layout}
    \end{subfigure}
    
    \caption{\textbf{A variation graph and its visualization example} --- the three genomes are depicted in different colors; the path of interconnected nodes represents the original genome.}
    \label{fig:variation-graph-demo}
\end{figure}

A prevalent pangenomic model to represent these differences is the variation graph~\cite{variation_graphs}. 
As illustrated in Fig.~\ref{fig:variation-graph-demo}, the variation graph captures both genomic sequences and variations amongst them. 
The data structure of a variation graph, formed by merging identical segments from multiple genomes into a single node, is depicted in Fig.~\ref{fig:variation-graph-abstract}. 
Its visualization, as seen in Fig.~\ref{fig:variation-graph-layout}, reveals variants including insertions, deletions, and single nucleotide variants (SNVs). 
In general, visualization is an effective way to reveal structural differences between genomes and gain insights~\cite{pangenome-graphs}.

\begin{figure}[!htbp]
    \centering
    \includegraphics[width=\linewidth]{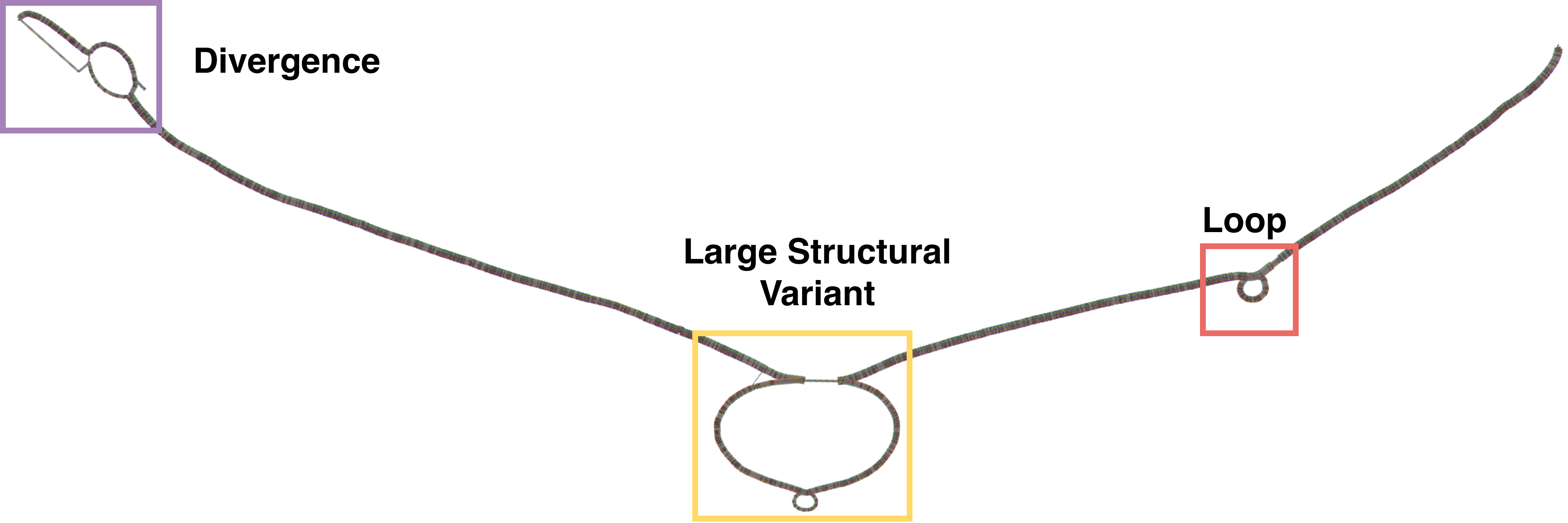}
    \caption{\textbf{Layout of the HLA-DRB1 gene} --- three distinct variant types are shown in the bounding boxes. }
    \label{fig:layout-DRB1}
\end{figure}

The visualization of a pangenome combines layout and rendering, with the layout component being fundamental to the graph visualization quality. 
In particular, the layout of a variation graph is crucial to variant discovery in large pangenomes.
Fig.~\ref{fig:layout-DRB1} shows the layout of the HLA-DRB1 gene~\cite{shiina2009hla}, which encodes an immune system protein associated with reduced severity of COVID-19 disease~\cite{HLA-DRB1-covid19}. 
Genome researchers can easily identify the location and structure of variants with an optimized, planar 2D layout of the variation graph, aiding in the study of pangenomes. 

However, general graph layout frameworks are not well-suited to effectively lay out pangenome graphs. 
This limitation primarily arises from the unique biological significance associated with the nodes and paths within a pangenome. 
Currently, only specialized tools~\cite{mikheenko2019AGB, Bandage, pangenome-odgi-layout-bioinformatics,libhandlegraph-garrison2018,gonnella2019gfaviz} for pangenome graphs offer effective support. 
Yet, pangenome graph layout, especially for large human pangenome graphs, remains an extremely time-consuming process. 
The current state-of-the-art approach \textit{odgi-layout}~\cite{pangenome-odgi-layout-bioinformatics} requires hour-scale time to generate the layout of the variation graph for a single human chromosome with a 32-core server-class Intel Xeon CPU. 
In addition, the layout process often requires multiple rounds of parameter tuning to achieve an optimal layout, resulting in a bottleneck in the pangenome analysis pipeline.

This work aims to accelerate the computation of pangenome graph layouts, a crucial step in pangenomics. We show that the pangenome layout algorithm exhibits a substantial degree of data-level parallelism, albeit underutilized in the current state-of-the-art CPU implementation~\cite{pangenome-odgi-layout-bioinformatics}, impeding progress in pangenome research. With significant data parallelism available, GPU acceleration holds promise for this application. However, challenges arise due to the irregular data access pattern and memory-bound nature of the algorithm.

In this paper, we present a novel solution to pangenome graph layout computation, by leveraging the computational power of modern GPUs and optimizing the data access pattern. Our approach not only accelerates layout computation but also improves the overall efficiency and scalability of pangenomics analyses.
Our main contributions are as follows:

\begin{itemize}[leftmargin=*]
    \item To our knowledge, we present the first GPU-based solution to accelerate pangenome graph layout, which enables minute-scale layout for the entire chromosome dataset. 
    Our implementation achieves an average speedup of 57.3$\times$ compared to an optimized, state-of-the-art CPU implementation.
    We will open-source our software in a format that facilitates easy integration into the pangenomic analysis pipeline. 

    \item To identify the performance bottleneck, we perform a detailed workload characterization of the pangenome graph layout algorithm. Our analyses indicate that this workload has a highly irregular data access pattern and is memory-bound. Thus, a na\"{\i}ve approach is inadequate for fully exploiting GPU's computational capabilities.

    \item We introduce three key optimizations to improve GPU performance: (1) optimizing the data layout for improved cache efficiency, (2) enabling coalesced memory accesses by coalescing random states, and (3) reducing warp divergence through warp merging. 

    \item We propose a quantitative metric called sampled path stress to assess the quality of GPU-generated layouts in a scalable manner. Through a case study, we demonstrate the potential to explore performance-quality trade-offs using this metric, leading to additional speedup. 

\end{itemize}

\section{Background}
\label{sec:background}
This section introduces the background of pangenomics, its variation graph representation, and its graph layout algorithm.

\subsection{Variation Graph}
\label{subsec:variation_graph}

Graph-based pangenomics aims to study genome variation within a population of samples. 
The variation graph serves as the primary model to describe graph-based pangenomes.

A variation graph $G=(P,V,E)$ is a directed graph composed of a set of \textit{paths} $P$, \textit{nodes} $V$ and \textit{edges} $E$, as shown in Fig.~\ref{fig:variation-graph-abstract}. 
Each \textit{node} represents a nucleotide sequence, each \textit{edge} represents the connection of an ordered pair of nodes, and each \textit{path} describes a walk over nodes. 

The path consists of interconnected nodes and represents the original genome, e.g., path 2 in Fig. \ref{fig:variation-graph-abstract} embodies a genome sequence of \texttt{AATGC...TAGCAAAC}. 
While most nodes are shared across all paths, variants exist in the form of unique nodes. 
These variants are revealed by visualizing the variation graph, as shown in Fig.~\ref{fig:variation-graph-layout}. 
For instance, the \texttt{T} insertion in path 2 serves as a variant and is the primary discovery focus. 


Variation graphs representing biological sequences typically exhibit a linear structure, as opposed to the more commonly encountered planar graphs. 
This characteristic stems from the linear nature of the genome sequences they represent, where the majority of segments are identical due to sequence homology. 
Consequently, variation graphs display a notably low average node degree and density. 
As an example, the average node degree of human pangenome graphs released by the HPRC~\cite{draft_human_pangenome_reference} is $1.4$, and the average density is $3.5\times10^{-7}$. 
These graph properties, along with the genome-specific path information, make variation graphs particularly unique, opening opportunities for ad-hoc algorithmic optimizations. 

\subsection{Pangenome Graph Layout}
\label{subsec:graph-layout}


The aim of a pangenome graph layout is to organize nodes and edges in order to highlight the genetic variation present in the genomes represented in the graph. This enables the large-scale study of the diversity and evolution embodied in tens or hundreds of genomes. For example, the layout structure of a pangenome graph representing the 5 acrocentric human chromosomes of the HPRC pangenome revealed heterologous recombination in the human pangenome~\cite{Guarracino2023}.

Existing general graph layout frameworks~\cite{gephi,graphdraw-by-SGD} struggle to reveal the structural variants of pangenome graphs.
We illustrate this by using Gephi~\cite{gephi} to lay out the HLA-DRB1 gene with algorithms including Fruchterman-Reingold~\cite{fruchterman1991graph}, ForceAtlas2~\cite{jacomy2014forceatlas2} and Yifan Hu~\cite{hu2005efficient}. 
These algorithms, while creating 2D structures, fail to uncover the underlying structural variants. 
This is due to their design for calculating distances between all nodes, whereas pangenome graphs only consider nodes on the same path meaningful.

Given that both the biological meanings of nodes and paths must be factored into the layout process, only specialized tools for pangenome graphs prove effective. 
Among these, the current state-of-the-art approach is \textit{odgi-layout}, which is part of the comprehensive pangenome analysis framework ODGI~\cite{ODGI}. 
By adapting Zheng et al.'s work~\cite{graphdraw-by-SGD} to the pangenomic field, \textit{odgi-layout} utilizes a path-guided stochastic gradient descent (Path-Guided SGD) algorithm to minimize stress, a proxy metric quantifying the difference between reference and layout distances. 
With its multi-threaded CPU implementation, \textit{odgi-layout} stands as the only tool capable of handling whole-chromosome graphs with millions of nodes.

However, more efficient graph layout solutions are needed to rapidly compute layouts of increasingly large and complex pangenomes.
Indeed, \textit{odgi-layout} demands hours on a 32-core Intel Xeon CPU to generate a pangenome graph layout for just one human chromosome. 
Specifically, computing the layout of the chromosome 1 (Chr.1) pangenome --- the largest chromosomal pangenome released by HPRC --- alone exceeds 2.5 hours. Completing the layouts for all 24 chromosomal pangenome graphs from HPRC sums up to a significant 28 hours. 
Notably, running the layout computation once takes up nearly a third of the entire pangenomics analysis pipeline~\cite{garrison2023building}'s duration. 
Given that multiple runs are often performed for optimization, the layout computation becomes an even more pronounced bottleneck.

This performance issue impedes the study of large and/or complex pangenome graphs because of the prolonged layout generation times. Importantly, a fast layout solution would facilitate interactive visualization, allowing on-the-fly exploration of specific loci, genomic regions, entire chromosomes, or even whole genomes. This would further pave the way for the development of next-generation pangenome browsers, unlocking the study of population-scale genetic variability. This motivates us to pursue substantial acceleration in pangenome graph layout generation.

\subsection{Path-Guided SGD Algorithm}
\label{subsec:path-guided-sgd-algorithm}

\begin{algorithm}[!htbp]
    \caption{Path-Guided Pangenome Graph Layout}
    \label{code:algorithm}
    \raggedright
    \textbf{Input:} pangenome graph $G = (P,V,E)$, SGD schedule $S$, total iteration count $N_{iters}$ \\
    \textbf{Output:} a $2$D layout $\mathbf{L}$ consisting of line segments. $\mathbf{L}[n]$ returns an array of 2 vectors pointing to its endpoints given $n \in V$.
    \begin{algorithmic}[1]
        \State $N_{steps} \gets 10 \times \sum_{p \in P} |p|$ 
        \Comment{$|p|$: \# of nodes in path $p$}

        \For{$iter \gets 0$ to $N_{iters}$}
            \State $\eta \gets S[iter]$
            \Comment{learning rate}
            \For{$step \gets 0$ to $N_{steps}$} \textbf{in parallel}
                \State $p \gets \text{RandomSelect(}P\text{, } prob \propto |p| \text{)}$
                \State $cooling \gets (iter \geq N_{iters}/2) \text{ or FlipCoin()}$

                \If{$cooling$}
                    \State $n_i, n_j \gets \text{RandomSelect(}p\text{, Powerlaw)}$
                \Else
                    \State $n_i, n_j \gets \text{RandomSelect(}p\text{, Uniform)}$
                \EndIf
                \State $\mathbf{v_i} \gets$ FlipCoin() ? $\mathbf{L}[n_i]$.start : $\mathbf{L}[n_i]$.end
                \State $\mathbf{v_j} \gets$ FlipCoin() ? $\mathbf{L}[n_j]$.start : 
                $\mathbf{L}[n_j]$.end
                
                \State $stress_{ij} \gets ({(||\mathbf{v_i} - \mathbf{v_j}|| - d_{ref}})/{d_{ref}})^2$
                \Comment{loss}
                
                

                \State $(\mathbf{v_i}, \mathbf{v_j}) \gets (\mathbf{v_i}, \mathbf{v_j}) - \eta\nabla stress_{ij}$
                \Comment{update}

            \EndFor
        \EndFor
    \end{algorithmic}
\end{algorithm}

Alg.~\ref{code:algorithm} presents the pseudocode for the path-guided SGD algorithm used in \textit{odgi-layout}. 
This algorithm iteratively selects one pair of \textit{nodes} $(n_i, n_j)$ from the same path $p$ (lines 7-11). For each of these nodes, represented by a line segment in the layout, a \textit{visualization point} is selected (lines 12, 13). This yields a pair of \textit{visualization points} $(\mathbf{v_i},\mathbf{v_j})$, each corresponding to an endpoint of the respective node's line segment. 
This pair of \textit{visualization points} forms a loss function (known as stress) with its reference distance $d_{ref}$ and current layout distance $||\mathbf{v_i} - \mathbf{v_j}||$ (line 14). 
Then the coordinates are updated based on the gradient (line 15). 

\begin{figure}[!htbp]
    \centering
    \includegraphics[width=0.8\linewidth]{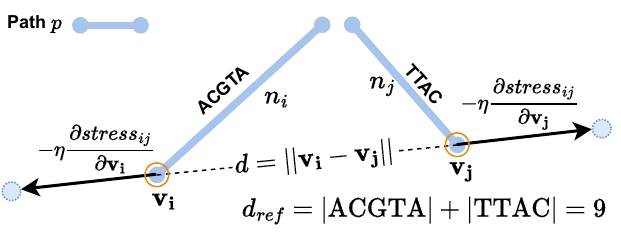}
    \caption{\textbf{Layout update within one step} --- $n_i$ and $n_j$ are two nodes representing the nucleotide sequences ``ACGTA'' and ``TTAC'', respectively. }
    \label{fig:pg-sgd-algorithm}
\end{figure}

The update process is illustrated in Fig.~\ref{fig:pg-sgd-algorithm}, 
where both nodes are moved against the direction of the gradient~\cite{graphdraw-by-SGD}. 

\section{Workload Characterization}
\label{sec:workload_characterization}

In this section, we describe the workload characterization of the multi-threaded CPU implementation of \textit{odgi-layout} on a 32-core Intel Xeon Gold 6246R 3.4GHz CPU. 
For detailed profiling, we use Linux Perf~\cite{perf} and Intel VTune profiler~\cite{intel-vtune-profiler}. 
We evaluate the layout computation on three representative pangenomes of varying sizes, as detailed in Table~\ref{tab:workload-datasets}.

\begin{table}[!htbp]
    \centering
    \caption{\textbf{Properties of representative pangenomes} --- \#~Nuc. is the number of nucleotides.}
    \label{tab:workload-datasets}
    \begin{tabular}{l|rrrrr}
        \toprule
        \multicolumn{1}{c|}{Pangenome} & \multicolumn{1}{c}{\# Nuc.} & \multicolumn{1}{c}{\# Nodes} & \multicolumn{1}{c}{\# Edges} & \multicolumn{1}{c}{\# Paths}  \\
        \midrule
        HLA-DRB1 & $2.2\times10^4$ &    $5.0\times10^3$	& $6.8\times10^3$	& $12$       \\
        MHC      & $5.9\times10^6$	&   $2.3\times10^5$	& $3.2\times10^5$	& $99$	\\
        Chr.1   & $1.1\times10^9$ &    $1.1\times10^7$	& $1.5\times10^7$	& $2,262$	\\
        \bottomrule
    \end{tabular}
\end{table}

Our analysis highlights three key observations: (1) the algorithm exhibits high data-level parallelism; (2) it is highly memory-bound; (3) randomness is critical to the layout quality. In the following, we delve deeper into each of them.

\subsection{Data-level Parallelism}
\label{subsec:parallelism}
The multi-threaded CPU implementation of \textit{odgi-layout} runs the inner loop in parallel and updates the layout asynchronously in a Hogwild!~\cite{hogwild} manner. 
This means that the \texttt{for} loop at line 4 in Alg.~\ref{code:algorithm} has high data-level parallelism for a graph with a large number of nodes. 
While the intrinsic race condition between parallel threads could introduce errors,
the layout quality is barely affected since pangenome graphs are so sparse that the probability of multiple threads updating the same nodes simultaneously is low. 

\begin{figure}[!htbp]
    \centering
    \begin{subfigure}[b]{0.325\linewidth}
        \centering
        \includegraphics[width=\linewidth]{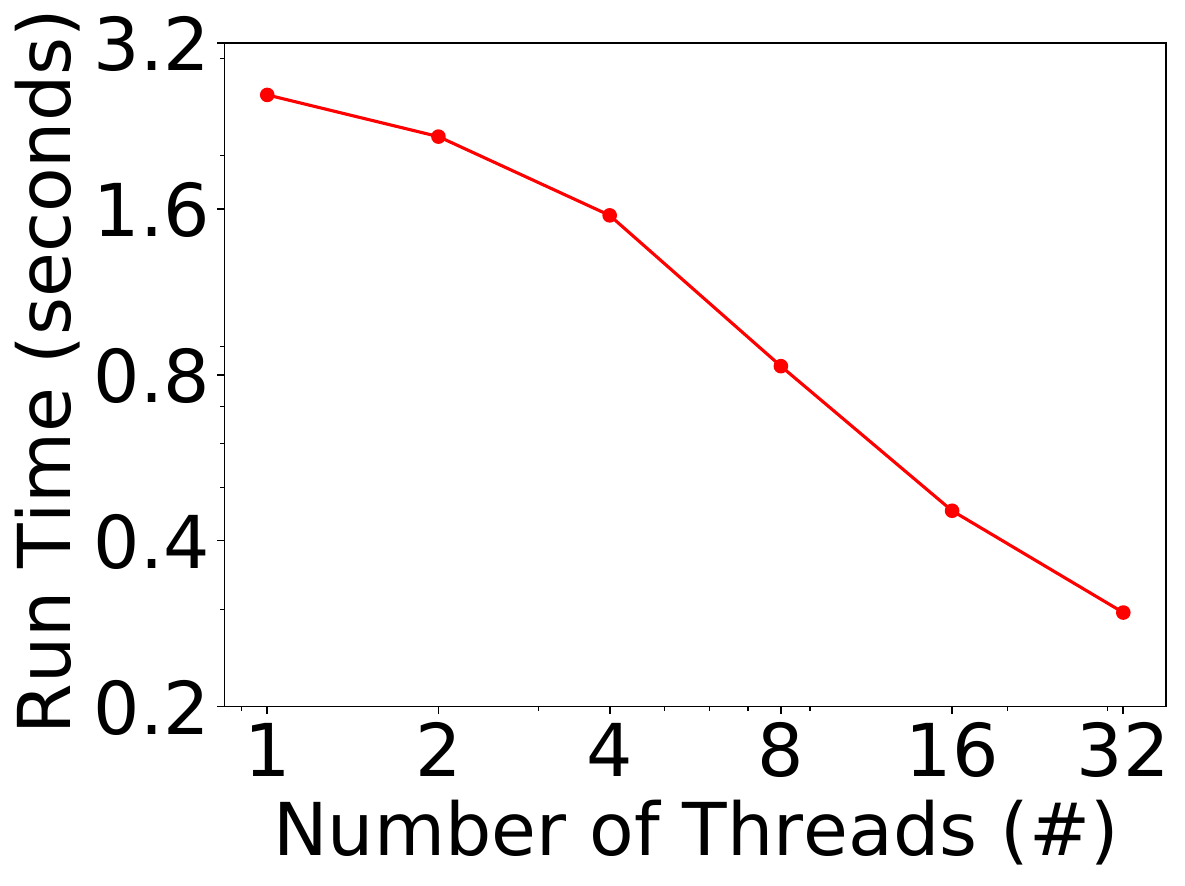}
        \caption{HLA-DRB1.}
    \end{subfigure}
    \centering
    \begin{subfigure}[b]{0.325\linewidth}
        \centering
        \includegraphics[width=\linewidth]{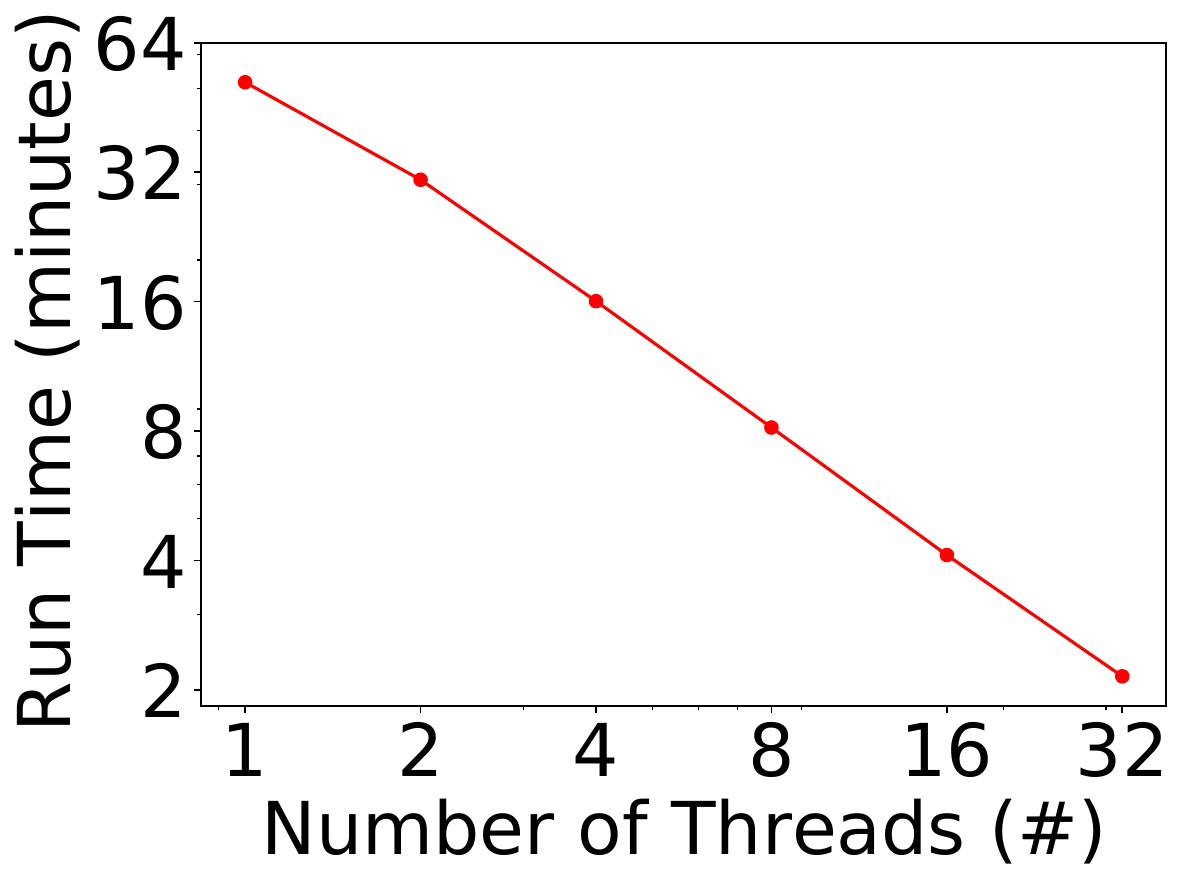}
        \caption{MHC.}
    \end{subfigure}
    \centering
    \begin{subfigure}[b]{0.325\linewidth}
        \centering
        \includegraphics[width=\linewidth]{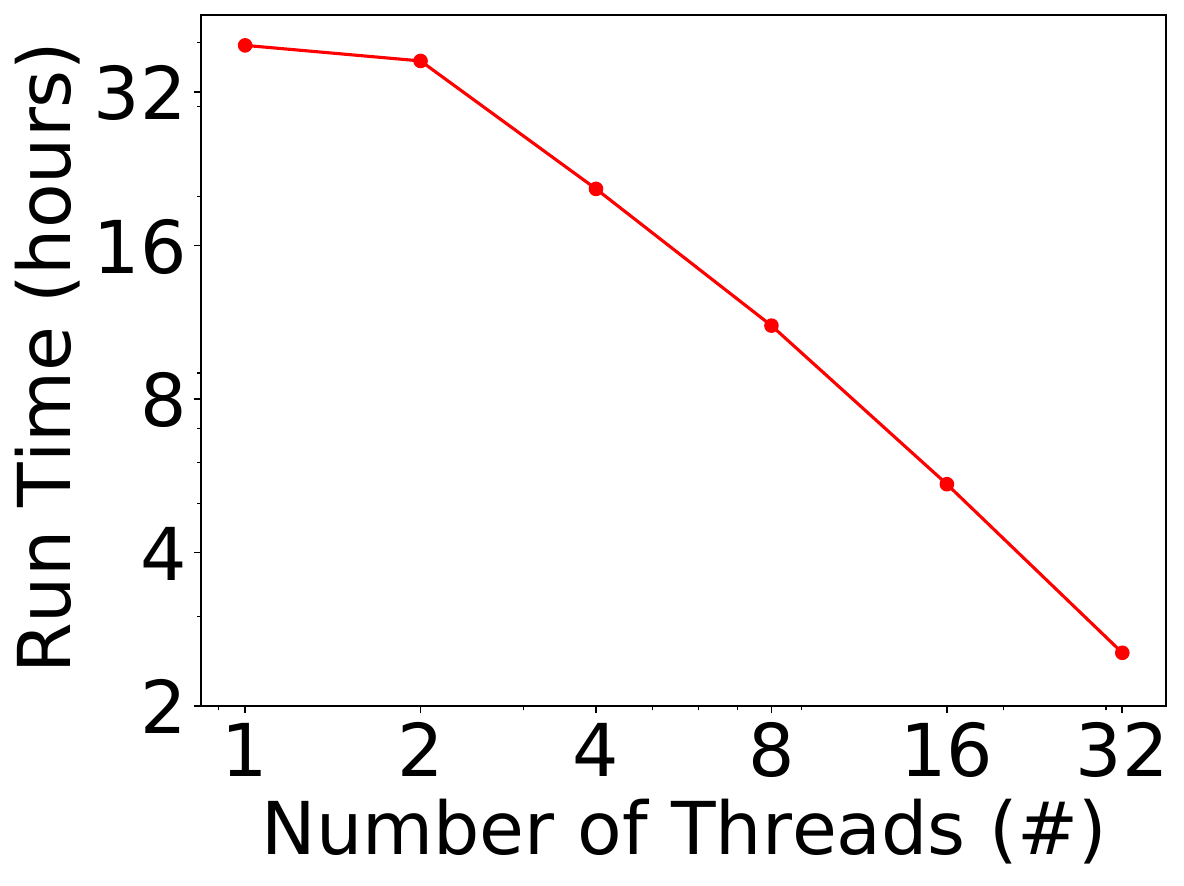}
        \caption{Chr.1.}
    \end{subfigure}
    
    \caption{\textbf{Scaling of \textit{odgi-layout}.}}
    \label{fig:workload-cpu-scale}
\end{figure}

Fig.~\ref{fig:workload-cpu-scale} reveals a linear scaling pattern of \textit{odgi-layout} with CPU threads. 
However, the CPU cannot fully take advantage of the high degree of data-level parallelism that exists in the inner loop, particularly for larger graphs. 
For instance, the pangenome graph of the human chromosome 1 (Chr.1) requires six billion node pair updates per iteration, making it less ideal for a CPU with a limited number of threads.

\subsection{Memory-Bound}
\label{subsec:memory-bound}

\begin{figure}[!htbp]
    \centering

    \centering
    \begin{subfigure}[b]{0.325\linewidth}
        \centering
        \includegraphics[width=\linewidth]{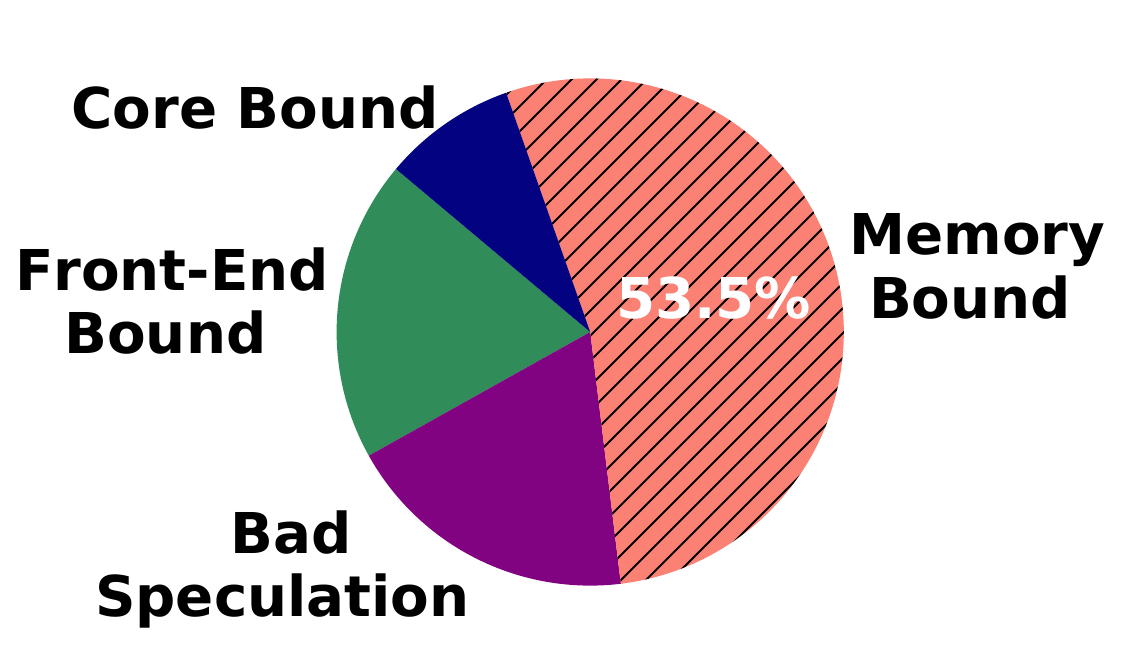}
        \caption{HLA-DRB1.}
    \end{subfigure}
    \centering
    \begin{subfigure}[b]{0.325\linewidth}
        \centering
        \includegraphics[width=\linewidth]{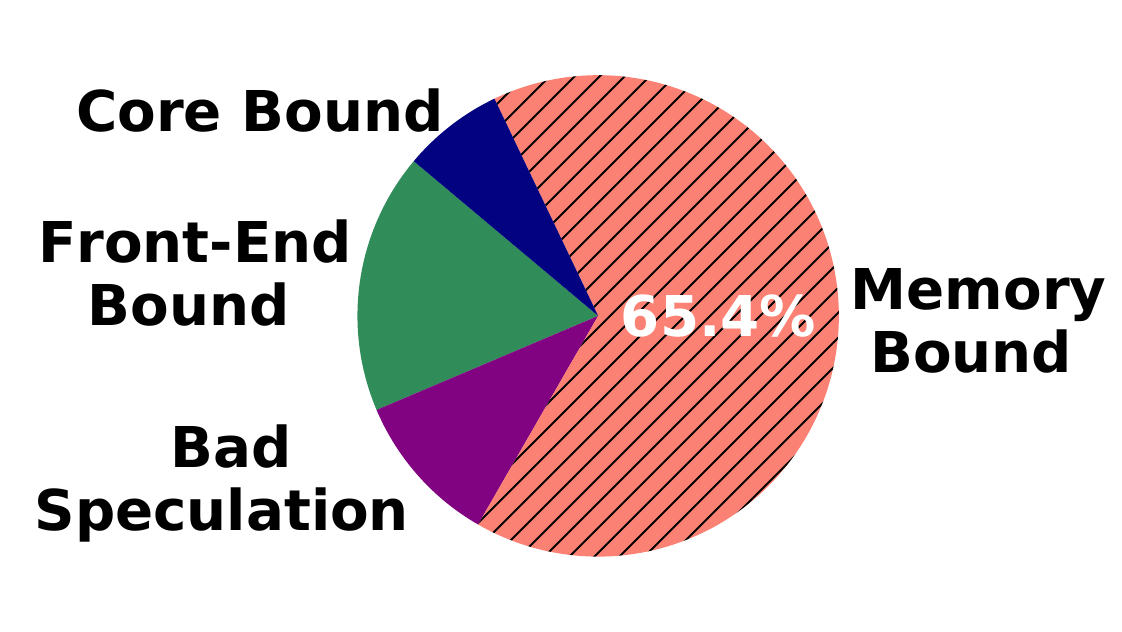}
        \caption{MHC.}
    \end{subfigure}
    \centering
    \begin{subfigure}[b]{0.325\linewidth}
        \centering
        \includegraphics[width=\linewidth]{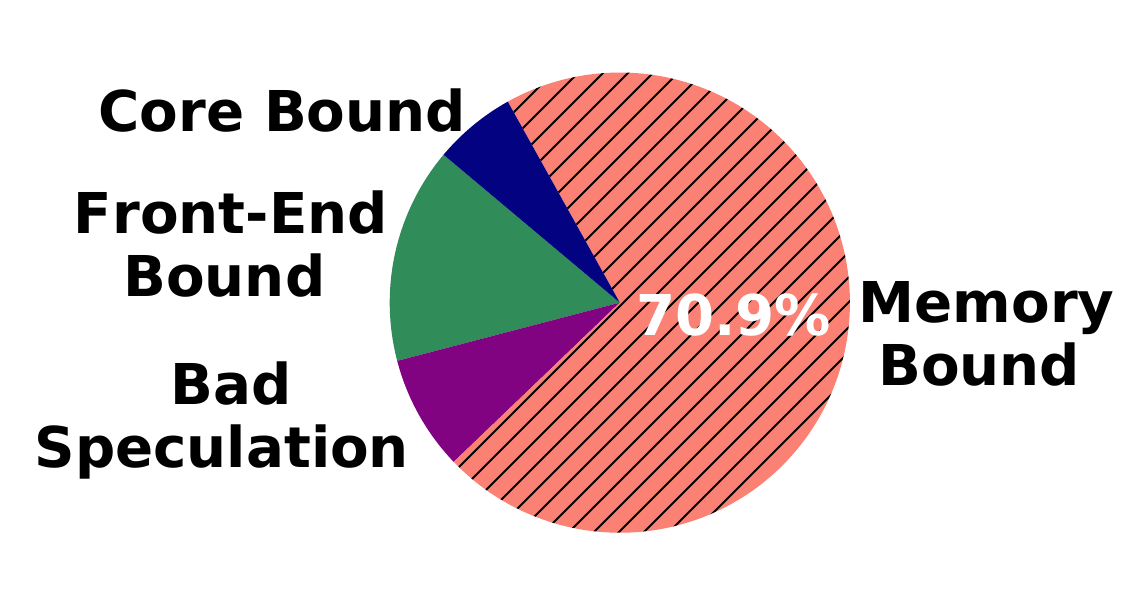}
        \caption{Chr.1.}
    \end{subfigure}
    \caption{\textbf{Microarchitecture bottleneck analysis with VTune.}}
    \label{fig:workload-microarchitecture-memory-bound}
\end{figure}

We use the top-down approach proposed in~\cite{yasin2014-topdown-approach} to identify the performance bottleneck. Fig.~\ref{fig:workload-microarchitecture-memory-bound} displays the results of the bottleneck analysis. It is apparent that \textit{odgi-layout} uses a significant portion of the microarchitecture's pipeline slots for memory operations on all three graphs, demonstrating its memory-bound nature.
We then profile the memory stall and cache performance of \textit{odgi-layout}. As illustrated in Table~\ref{tab:memory-bound-chr1}, workload performance is bottlenecked by a high percentage of memory stall cycles and a significant miss rate of last-level cache (LLC) loads. As a result, the memory operations dominate the run time of the layout process.

\vspace{-1em}

\begin{table}[!htbp]
    \centering
    \caption{\textbf{Memory stall and cache performance of \textit{odgi-layout} profiled by Perf. }}
    \label{tab:memory-bound-chr1}
    \begin{tabular}{l|rrr}
        \toprule
        Pangenome                     & \multicolumn{1}{c}{HLA-DRB1} & \multicolumn{1}{c}{MHC} & \multicolumn{1}{c}{Chr.1} \\
        \midrule
        Run Time (h:mm:ss)            & 0:00:00.4  & 0:01:47    & 2:32:38     \\
        Memory Stall Cycle Percentage & $67.67\%$  & $78.07\%$ & $77.38\%$ \\
        LLC-load Miss Rate            & $75.09\%$  & $77.84\%$ & $89.88\%$   \\
        \bottomrule
    \end{tabular}
\end{table}

We observe that the random memory accesses to $\mathbf{L}$ (lines 12, 13) and obtaining $d_{ref}$ outweigh the computational part (lines 14, 15).
Given the massive size of these data structures, e.g., the graph of Chr.1 is composed of 11.1M nodes, the scope for data reuse is severely limited due to random memory access. 
This leads to the unusually high LLC load miss rate.

Additionally, the repeated use of pseudo-random number generator (PRNG) (lines 5, 6, 8, 10, 12, 13) increases memory traffic. \textit{odgi-layout} uses Xoshiro256+~\cite{blackman2021xoshiro}, a PRNG utilizing linear-feedback shift registers (LFSR). 
LFSR-based PRNG is known for its low computational requirements, which adds to the memory-bound nature of the layout process.

\subsection{Randomness \& Layout Quality}
\label{subsec:randomness_quality}

\begin{figure}[!htbp]
    \centering
    \includegraphics[width=\linewidth]{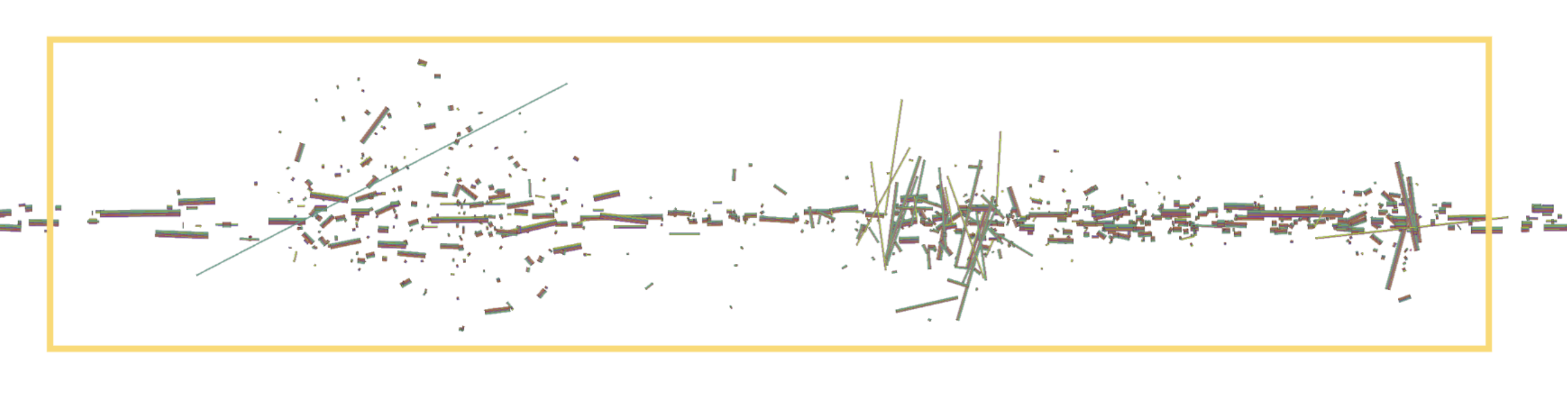}
    \caption{\textbf{Layout of poor quality} --- the yellow box captures the ``Large Structural Variant'' region in Fig.~\ref{fig:layout-DRB1}.}
    \label{fig:workload-bad-layout-less-randomness}
\end{figure}

Randomness is essential for fast convergence and high-quality layout generation in this path-guided SGD algorithm. 
This is consistent with the discussion in the paper by Zheng et al.~\cite{graphdraw-by-SGD}, from which the path-guided SGD algorithm was adopted.
Random path and node pair selections (lines 5, 8, 10) are performed in each step to ensure the layout quality, as a na\"{\i}ve iteration could cause the algorithm to get stuck in local minima due to biases.
Fig.~\ref{fig:workload-bad-layout-less-randomness} shows a non-converged layout created by forcing all selected pairs of nodes to be 10 hops away. 
This node pair selection scheme significantly reduces randomness in node selection and does not converge within the same number of iterations. 
In contrast, the optimized layout of the same gene shown in Fig.~\ref{fig:layout-DRB1} clearly reveals the variants, which are the primary targets of pangenome graph layout. 
\vspace{-1em}

\section{Pangenome Graph Layout in PyTorch}
\label{subsec:pytorch-design}

As previously discussed, CPUs cannot fully exploit the substantial data-level parallelism within the pangenome graph layout algorithm.
Since the algorithm relies on SGD-based optimization, adopting PyTorch~\cite{pytorch}, a deep learning framework optimized for gradient computation, is an attractive option for implementing the layout algorithm on GPUs.

In this section, we introduce a PyTorch-based implementation of the algorithm and assess its performance on the MHC pangenome graph with an NVIDIA RTX A6000 GPU. We employ NVIDIA Nsight Systems~\cite{nvidia-nsight-systems} for detailed profiling. Our analysis not only reveals the limitations of a basic PyTorch implementation but also underscores the challenges in achieving effective GPU acceleration of the pangenome layout.

\subsection{Implementation and Performance Analysis}
\label{subsec:pytorch-performance-analysis}

We utilize PyTorch to solve the layout optimization problem following the neural network training procedure --- Each data instance is a node pair $(n_i,n_j)$, with its ground-truth label as $d_{ref}$.  The layout coordinates $\mathbf{L}$ act as the adjustable weights that are updated in each step based on the gradient of the stress function. We process a batch of node pairs simultaneously to leverage the data-level parallelism of the algorithm. 

The performance of the PyTorch implementation on the MHC pangenome is shown in Table~\ref{tab:workload-torch-baseline-perf}, where we measure the GPU run time and compare it to the 32-thread CPU baseline, which completes in 107 seconds.
The run time decreases as the batch size increases, up to a batch size of 1 million. Beyond this point, there is no further linear scaling.

We also assess the layout quality to understand the impact of larger batch sizes by visual inspection. 
As mentioned in Sec.~\ref{subsec:parallelism}, excessive asynchronous updates by too many threads could reduce the effectiveness of these updates, resulting in layout quality degradation. 
This is reflected in the increasing node stress with larger batch sizes.  
By visual inspection, the design with a batch size of 10M has some layout quality degradation, and the design with a batch size of 100M does not converge to a valid layout.

\begin{table}[!htbp]
    \centering
    \caption{\textbf{Performance of the PyTorch implementation} ---  the speedup is compared to the 32-thread CPU baseline. }
    \label{tab:workload-torch-baseline-perf}
    \resizebox{0.8\columnwidth}{!}{
    \begin{tabular}{l|rrrr}
        \toprule
        Batch Size        & Run Time (s)   & Speedup  & Quality  \\
        \midrule
        10K        & 702.2         & 0.2x           & Good \\
        100K       & 67.3          & 1.6x           & Good \\
        1M         & 15.6          & \textbf{6.8x}  & Good \\
        10M        & 14.3          & 7.5x           & Satisfying  \\
        100M       & 11.8          & 9.1x           & Poor \\
        \bottomrule
    \end{tabular}
    }
\end{table}

The PyTorch implementation achieves a 6.8$\times$ speedup over the CPU baseline on MHC.
However, this approach does not fully exploit the potential of the GPU due to the lack of tailored optimizations for the memory-bound nature of the application and the GPU architecture. 

\begin{figure}[!htbp]
    \centering
    \begin{subfigure}[b]{0.32\linewidth}
        \centering
        \includegraphics[width=\linewidth]{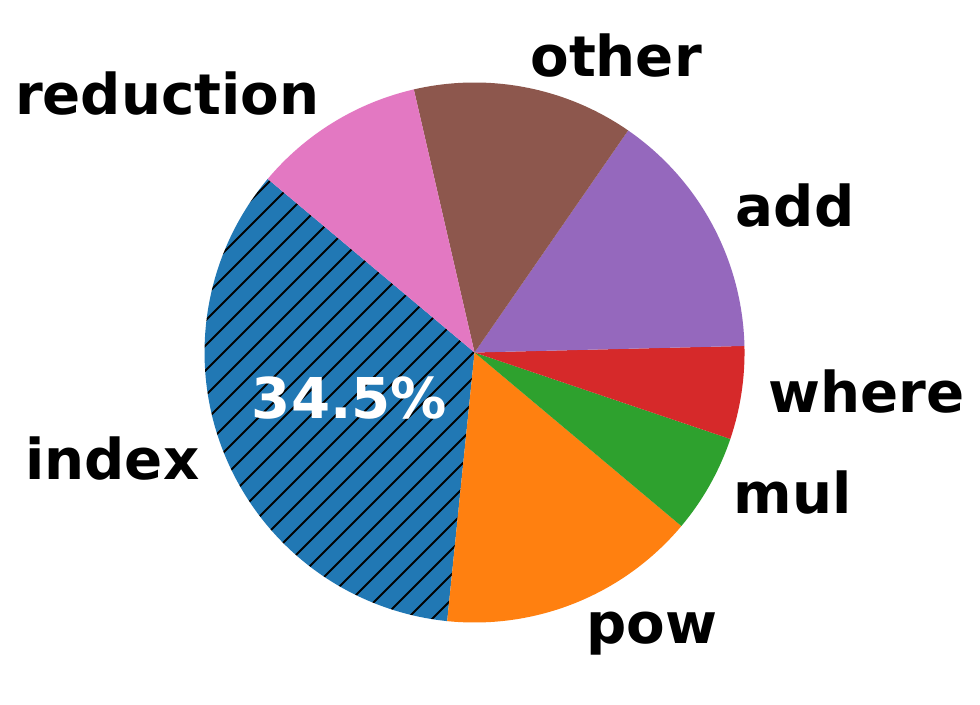}
        \caption{Batch size = 100K.}
        \label{fig:workload-torch-piechart-100K}
    \end{subfigure}
    \centering
    \begin{subfigure}[b]{0.32\linewidth}
        \centering
        \includegraphics[width=\linewidth]{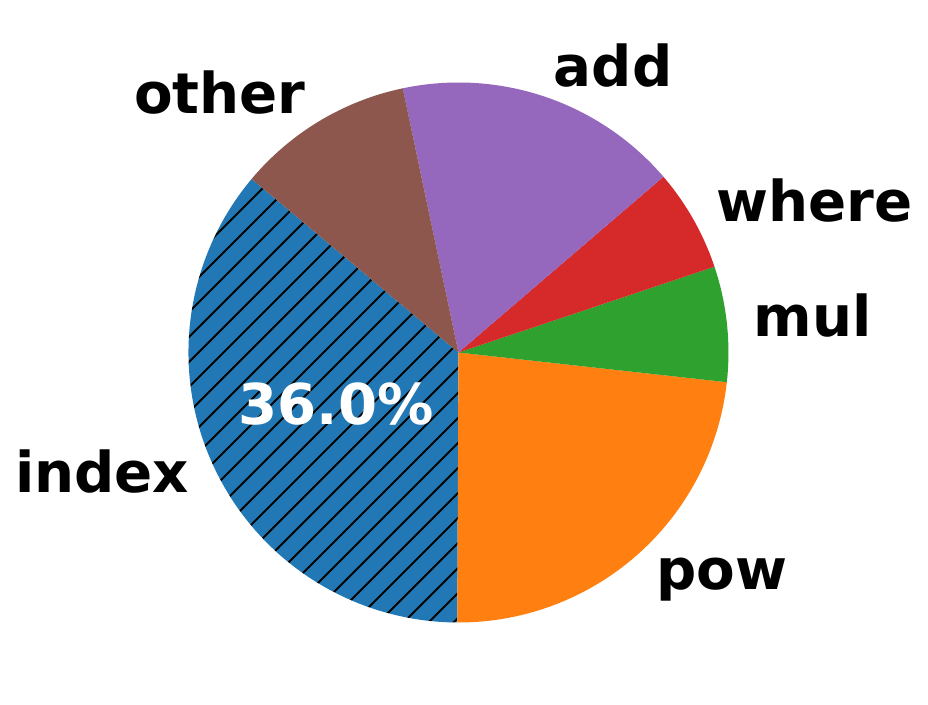}
        \caption{Batch size = 1M.}
        \label{fig:workload-torch-piechart-1M}
    \end{subfigure}
    \centering
    \begin{subfigure}[b]{0.32\linewidth}
        \centering
        \includegraphics[width=\linewidth]{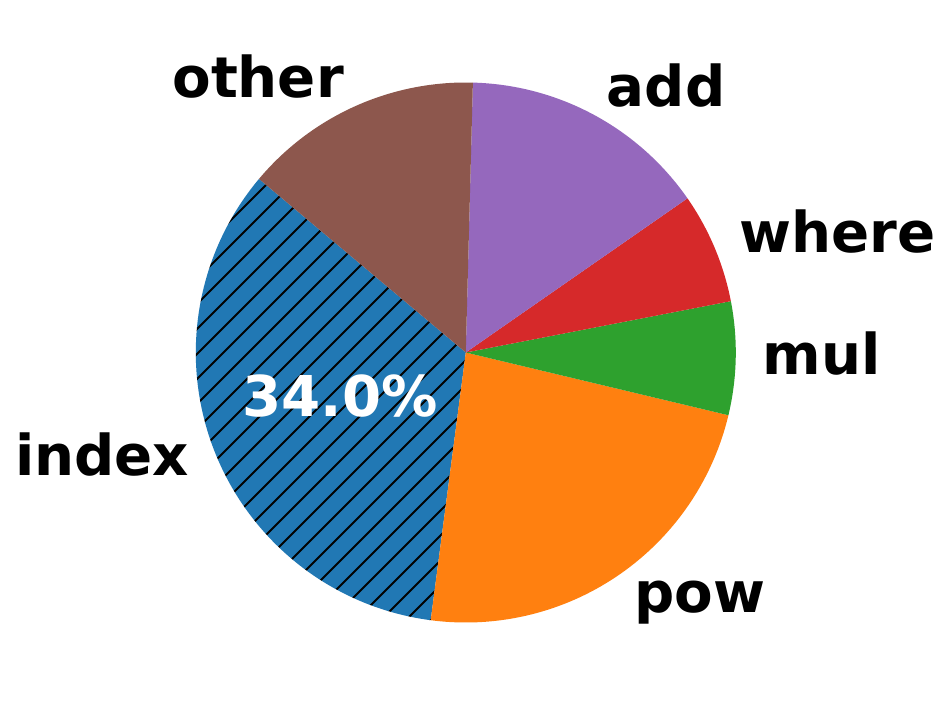}
        \caption{Batch size = 10M.}
        \label{fig:workload-torch-piechart-10M}
    \end{subfigure}

    \caption{\textbf{Kernel time breakdown of the PyTorch implementation, profiled by NVIDIA Nsight Systems} --- only kernels accounting for over 2\% of total GPU time are included. The shaded \texttt{index} is the memory operation.}
    \label{fig:workload-torch-piechart}
\end{figure}

Fig.~\ref{fig:workload-torch-piechart} presents the breakdown of kernel time for the PyTorch implementation when using different batch sizes, demonstrating that the indexing kernel consumes the most time. 
Taking into account the profiling results in Sec.~\ref{subsec:memory-bound}, it is evident that memory operations are the primary time-consuming operations on both the CPU and GPU. 
Given the algorithm's inherent randomness leading to a random memory access pattern, combined with this memory operation dominance, an effective data layout is crucial to enhance performance on both hardware platforms. 
However, neither the CPU baseline nor the current PyTorch implementation has a customized data layout. 

Another challenge arises from PyTorch's tensor-based programming model. The implementation groups multiple node pairs into long tensors for computation and memory operations. Due to the large number of node pairs, multiple batches are needed per iteration, resulting in numerous CUDA kernel launches, as shown in Table~\ref{tab:workload-cuda-kernel}. This leads to significant overhead in kernel launches and unnecessary implicit synchronization between kernels, which is not needed for this specific application that permits asynchronous Hogwild! style updates.

\begin{table}[!htbp]
    \centering
    \caption{\textbf{CUDA kernel launching overhead.}}
    \label{tab:workload-cuda-kernel}
    \begin{tabular}{l|rrr}
        \toprule
        Batch Size                    & 100K   & 1M & 10M   \\
        CUDA kernels launched (\#)    &  6,562,860  &    651,480     & 64,080    \\
        Time percentage of CUDA API   & $76.4\%$  & $20.2\%$  & $2.1\%$      \\
        \bottomrule
    \end{tabular}
\end{table}

Furthermore, using PyTorch, a high-level framework, makes it challenging to implement low-level, customized optimizations tailored to the GPU architecture. 
The highly-optimized kernels that PyTorch relies on come from its backend libraries. 
These are fixed and not tailored for our specific workload, which means that issues like conditional branching (lines 7, 9 in Alg.~\ref{code:algorithm}) and uncoalesced memory access can still significantly impair GPU performance.

\subsection{Challenges to Efficient GPU Offloading}
\label{subsec:challenges-in-gpu}

By characterizing the pangenome graph layout workload and implementing a basic PyTorch implementation, 
we have identified several challenges that must be addressed in order to fully leverage the power of GPUs.

\begin{itemize}[leftmargin=*]
    \item Numerous CUDA kernels launched by PyTorch lead to a notable overhead due to redundant memory operations and synchronization. This is addressed in Sec.~\ref{subsec:design-base-cuda-kernel}.

    \item The application is memory-bound on both CPUs and GPUs. Dominant memory operations and irregular access patterns necessitate an effective data layout to minimize memory traffic. This is addressed in Sec.~\ref{subsec:design-cache-friendly-ds}.

    \item The conditional branching and uncoalesced memory access can degrade GPU performance. This is addressed in Sec.~\ref{subsec:design-coalesced-random-state} and Sec.~\ref{subsec:design-warp-merging}. 

\end{itemize}

\section{Optimized GPU Implementation}
\label{sec:design}

In this section, we describe our GPU design with customized optimizations to address the challenges highlighted in Sec.~\ref{subsec:challenges-in-gpu}. 
First, we introduce a base CUDA kernel for pangenome graph layout to exploit the high degree of data-level parallelism.
Then, we detail three optimization techniques: a \textit{cache-friendly data layout} for the pangenome graph, 
\textit{coalesced random states}, and \textit{warp merging}.

\subsection{CUDA Kernel for Pangenome Graph Layout}
\label{subsec:design-base-cuda-kernel}

\begin{figure}[!htbp]
    \centering
    \includegraphics[width=0.85\linewidth]{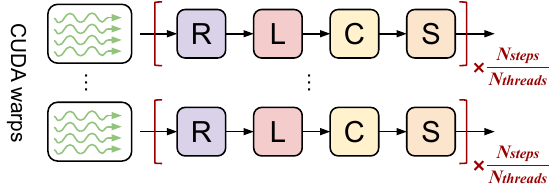}
    \caption{\textbf{CUDA kernel execution} --- one update step includes pseudo random number generation (R), node pair loading (L), computing the updated value (C), and storing the result (S). }
    \label{fig:design-gpu-parallelism}
\end{figure}

Our base CUDA kernel design for pangenome graph layout is shown in Fig.~\ref{fig:design-gpu-parallelism}. Each GPU thread runs the update steps (lines 4-16 in Alg.~\ref{code:algorithm}) in parallel. 
Within a single CUDA kernel launch, all GPU threads collectively contribute to completing the $N_{steps}$ steps required per iteration. 

The memory-bound nature of the algorithm would lead to frequent memory stalls. When a warp is stalled, the GPU warp scheduler attempts to switch to another available warp to hide memory latency. 
The abundant data-level parallelism in our design ensures the amount of available warps, thereby improving streaming multiprocessors (SM) utilization.


In our method, a single CUDA kernel is launched per iteration, with inter-block synchronization occurring only after all steps in an iteration are completed. 
Therefore, with the default setting of $N_{iters}$ of 30, a total of 31 CUDA kernels are launched, including one additional kernel launch for initialization. 
This achieves implicit kernel fusion compared to our preliminary PyTorch implementation, which greatly reduces the overhead due to the numerous CUDA kernels launched, as discussed in Table~\ref{tab:workload-cuda-kernel}.


Here, we also build a lean data structure specifically for the pangenome graph layout application. 
As a part of the comprehensive pangenome analysis framework ODGI, the current SOTA \textit{odgi-layout} uses the ODGI data structure. 
Therefore, the data structure includes numerous fields, some of which are not relevant to \textit{odgi-layout}, resulting in a suboptimal data structure for pangenome graph layout.

The lean data structure in our CUDA kernel retains only the data fields used in the pangenome graph layout process.
For instance, the ODGI data structure represents the nucleotide sequence as a string, and invoking the \texttt{.size()} method returns the size; our lean data structure directly stores the sequence length since the content of the string is not used in the pangenome graph layout. 
Note that this lean data structure can be easily transformed from the ODGI data structure, leading to an easy integration into the ODGI framework.

Since \textit{odgi-layout} and the external libraries used \cite{libhandlegraph-garrison2018} were developed for the multi-core CPU, the data structures are heavily dependent on the use of dynamic containers such as vectors.
The GPU provides limited support for dynamic data constructs, so we manually implement the necessary data structures and functions in our CUDA kernel.

\subsection{Kernel Optimizations}
\label{subsec:design-optimized-cuda-kernel}

To address the memory-bound nature of the application, we introduce three kernel optimization methods: a \textit{cache-friendly data layout} for pangenome graphs to improve cache locality, \textit{coalesced random states} to enable coalesced memory accesses, and \textit{warp merging} to reduce warp divergence.

\subsubsection{Cache-friendly Data Layout}
\label{subsec:design-cache-friendly-ds}

Our data structure for the pangenome graph layout includes node data and path data.
The node data includes the sequence length of each node and the coordinates of the start and end points of the visualization, while the path data consists of the node ID, path ID, position, and orientation of each node as it traverses the paths.

ODGI maintains its core data structure for pangenome graph and develops auxiliary structures for the tools built upon it. 
For instance, the x and y coordinates, which are used exclusively in \textit{odgi-layout}, are organized into two arrays separate from the primary graph data structure. 
The data structure of our base CUDA kernel follows this design, resulting in a struct-of-arrays (SoA) layout.
This has a negative impact on cache performance for the pangenome graph layout workload.
To solve this problem, we propose a cache-friendly data layout by repacking data to match the memory access pattern of Alg.~\ref{code:algorithm}.

\begin{figure}[!htbp]
    \centering
    \begin{subfigure}[b]{\linewidth}
        \centering
        \includegraphics[width=\linewidth]{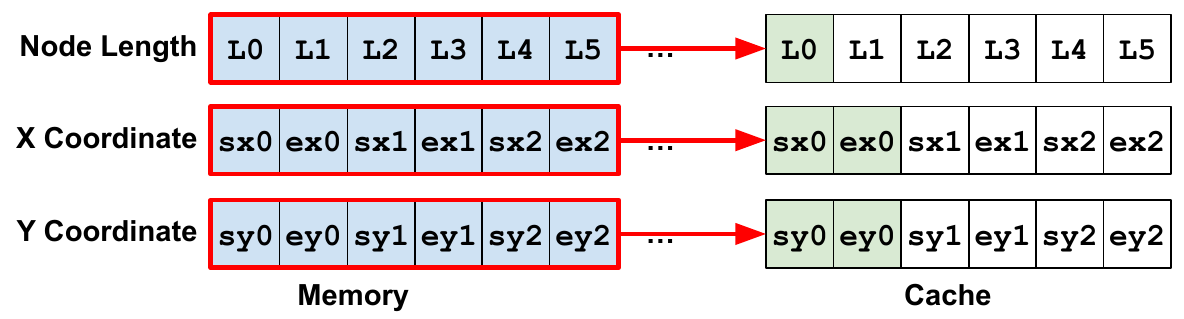}
        \caption{Original data layout. Every node incurs three memory accesses; the majority of cached data are not used due to randomness.}
        \label{fig:data-layout-comp-original}
    \end{subfigure}

    \begin{subfigure}[b]{\linewidth}
        \centering
        \includegraphics[width=\linewidth]{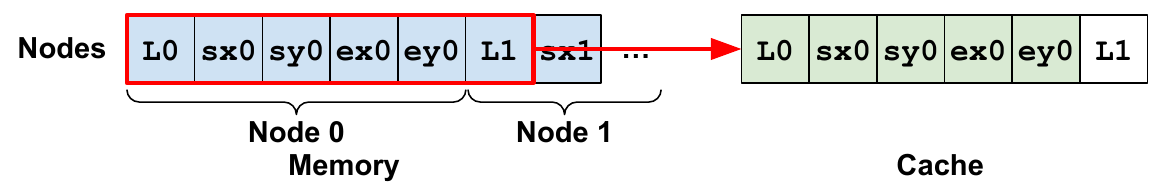}
        \caption{Cache-friendly data layout. One memory access for one node.}
        \label{fig:data-layout-comp-cache-friendly}
    \end{subfigure}

    \caption{\textbf{Cache-friendly data layout} --- \texttt{Li} is the length of node \texttt{i}; \texttt{sxi}, \texttt{syi}, \texttt{exi}, \texttt{eyi} are the x and y coordinates of the start and end points of the line segment for node \texttt{i}.}
    \label{fig:data-layout-comp}
\end{figure}

Fig.~\ref{fig:data-layout-comp} compares the proposed cache-friendly data layout with the original one in terms of access to node data during an update step.
When using the original data layout, updating a single node requires three separate memory accesses for three different arrays.
This is illustrated in Fig.~\ref{fig:data-layout-comp-original}. Although neighboring node data is cached, there is a high chance of eviction due to the random selection of node pairs.

In contrast, we use an array-of-structs (AoS) layout for node-related data, ensuring a cache-friendly design. Only one memory access is necessary for each node's data retrieval, as shown in Fig.~\ref{fig:data-layout-comp-cache-friendly}.
Since memory accesses to the start and end point coordinates are contiguous (lines 12, 13 in Alg.~\ref{code:algorithm}), this packing scheme improves spatial locality, thus removing traffic to higher-level caches and DRAM. 
The same principle applies to the path data which is not discussed in detail here.

\subsubsection{Coalesced Random States}
\label{subsec:design-coalesced-random-state}

Pseudo random number generator (PRNG) is heavily used in the algorithm. 
The CUDA cuRAND library~\cite{curand-library} utilizes the xorshift PRNG~\cite{xorshift-rngs}, a type of LFSR with low computational requirements.

To maintain layout quality, we map a set of random states to each SM, enabling each thread within a block to have its own random state. 
This ensures that threads generate uncorrelated random numbers, eliminating potential biases. 
However, this approach leads to a large number of memory accesses to the random states with concurrent running threads, which becomes the primary bottleneck for PRNG. 
As the GPU cache is shared by multiple warps running asynchronously, one warp's pangenome graph data may displace another warp's random states in the cache, increasing the risk of eviction.

\begin{figure}[!htbp]
\centering

    \begin{subfigure}[b]{0.5\linewidth}
        \centering
        \includegraphics[width=\linewidth]{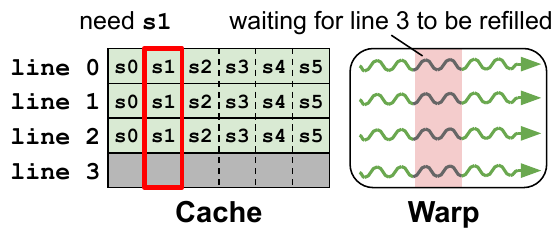}
        \caption{Original random states.}
        \label{fig:coalesced_rng-original}
    \end{subfigure}
    \begin{subfigure}[b]{0.45\linewidth}
        \centering
        \includegraphics[width=\linewidth]{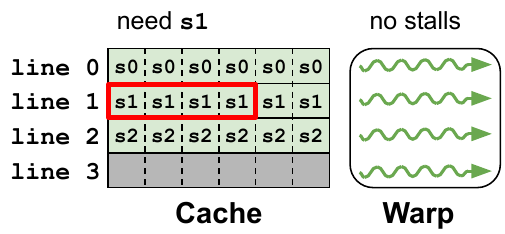}
        \caption{Coalesced random states.}
        \label{fig:coalesced_rng-coalesed}
    \end{subfigure}
    \caption{\textbf{Coalesced random states} --- in (a), a refill is required for any evicted cache line; in (b), a refill only happens when the warp accesses exactly the evicted cache line.}

    \label{fig:coalesced_rng}
\end{figure}

The cuRAND implementation represents each random state by a structure consisting of six 32-bit fields.
This object-oriented design forms an AoS data layout, with each thread having its own random state.
However, this data layout results in uncoalesced memory access to the random state, since the same field in different threads is not in contiguous memory.
Uncoalesced memory access to any random state requires much more frequent cache refills if some cache lines (e.g., cache line 3 in Fig.~\ref{fig:coalesced_rng-original}) are evicted.
This pattern amplifies global memory accesses and causes memory stalls. 

To solve this problem, we introduce a \textit{coalesced random states} method by transforming the AoS data layout into the SoA data layout.
As shown in Fig.~\ref{fig:coalesced_rng-coalesed}, this switch facilitates coalesced memory accesses to random states within a warp, storing the same field from multiple threads within the same cache line. 
In this way, a cache is only refilled from global memory when a warp requires an evicted cache line.

\subsubsection{Warp Merging}
\label{subsec:design-warp-merging}

The conditional branching (lines 7, 9 in Alg.~\ref{code:algorithm}) is crucial for generating a high quality pangenome graph layout.  
The non-cooling branch uniformly selects node pairs to create the coarse-grained layout, while the cooling branch selects node pairs at closer proximity with a power law distribution to refine the layout. 
However, this conditional branching structure leads to warp divergence. 
Since all 32 threads within a warp execute the same instruction, divergence forces some threads to idle, degrading GPU performance.

\begin{figure}[!htbp]
    \centering

    \includegraphics[width=0.8\linewidth]{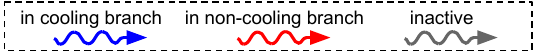}
    
    \vspace{3pt}
    \begin{subfigure}[t]{0.47\linewidth}
        \centering
        \includegraphics[width=\linewidth]{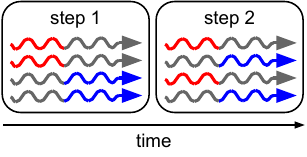}
        \caption{No warp merging. }
        \label{fig:no-warp-merging}
    \end{subfigure}    
    \begin{subfigure}[t]{0.35\linewidth}
        \centering
        \includegraphics[width=0.85\linewidth]{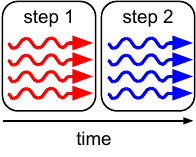}
        \caption{With warp merging. }
        \label{fig:warp-merging}
    \end{subfigure}
    \caption{\textbf{Warp merging} --- in (a), conditional branches cause warp divergence, leading to suboptimal thread utilization; in (b), all threads within a warp are active by selecting the same branch in an update step. }

    \label{fig:warp_optimization}
\end{figure}

To solve this problem, we introduce the warp merging method. 
As indicated in Fig.~\ref{fig:warp_optimization}, all threads within a warp select the same branch in an update step, keeping the threads constantly active.
This method is achieved by using a control thread within each warp to randomly select the branch.
The selection is then stored in shared memory, accessible to all threads within the same warp.

While warp merging causes threads within a single warp to select the same branch, resulting in reduced intra-warp randomness, the presence of multiple concurrently running warps on various SMs ensures different branches are chosen across warps. Consequently, the overall distribution of threads taking each branch remains consistent with the original algorithm, thereby preserving layout quality. 

\section{A Quantitative Metric for Pangenome Layouts}
\label{sec:metrics}

In our GPU implementation, we leverage a notably higher degree of data-level parallelism in comparison to the CPU baseline. As detailed in Section \ref{subsec:parallelism} and examined through experiments in Section \ref{subsec:pytorch-performance-analysis}, excessive parallelism may challenge the sparsity assumption underlying the Hogwild! asynchronous update, potentially compromising layout quality. Visual inspection, while useful, is subjective and not scalable since it relies on human evaluation of the results.  Consequently, there is a crucial need to quantify the quality of the GPU-generated layouts. 


In this section, we incorporate the stress metric, widely used in general graph layouts, into the pangenome graph to propose the \textit{path stress} with a GPU implementation, and then further apply sampling to solve the scalability issue.

\subsection{Path Stress}
\label{subsec:metrics-path-stress}

Prior studies~\cite{purchase1997aesthetic,gibson2013survey} have introduced various quantitative metrics to evaluate the aesthetic quality of general graph layouts, including stress, the number of edge crossings, the uniformity of edge lengths. 
However, each metric focuses on a single aspect, 
while some criteria contradict each other~\cite{haleem2019evaluating}. 
So far, there is no agreement on the most effective metric~\cite{dwyer2009comparison}, and the metric selection highly depends on which features of the graph you want to highlight in each use case~\cite{blythe1995effect,gibson2013survey}.

Therefore, since the pangenome graph layout algorithm is based on the popular energy-based algorithms by minimizing stress~\cite{graphdraw-by-SGD,kamada1989algorithm,gansner2005graph-stress-majorization}, we incorporate stress (line 14 in Alg.~\ref{code:algorithm}) with the unique path property of the pangenome graph, forming the \textit{path stress}, defined in Equation \ref{eq:path_stress}. 

\begin{equation}\label{eq:path_stress}
    \resizebox{0.8\linewidth}{!}{$\displaystyle
        path\_stress = \frac{
            \sum_{p \in P} \sum_{n_i,n_j \in p} 
            stress(n_i, n_j)
        }{
            N_{total\_node\_pairs}
        } 
    $}
\end{equation}

Here $stress(n_i, n_j)$ is the average stress of all four combinations of the start and end points of node $n_i$ and $n_j$.
The path stress is calculated by averaging the stress across all node pairs on all paths. 
The key distinction between path stress and the standard stress used for general graphs is that path stress only considers node pairs on the same path.
This aligns with the layout algorithm, as $d_{ref}$ only considers distances within the same path. 

\begin{figure}[!htbp]
    \centering
    \begin{subfigure}[t]{0.45\linewidth}
        \includegraphics[width=\linewidth]{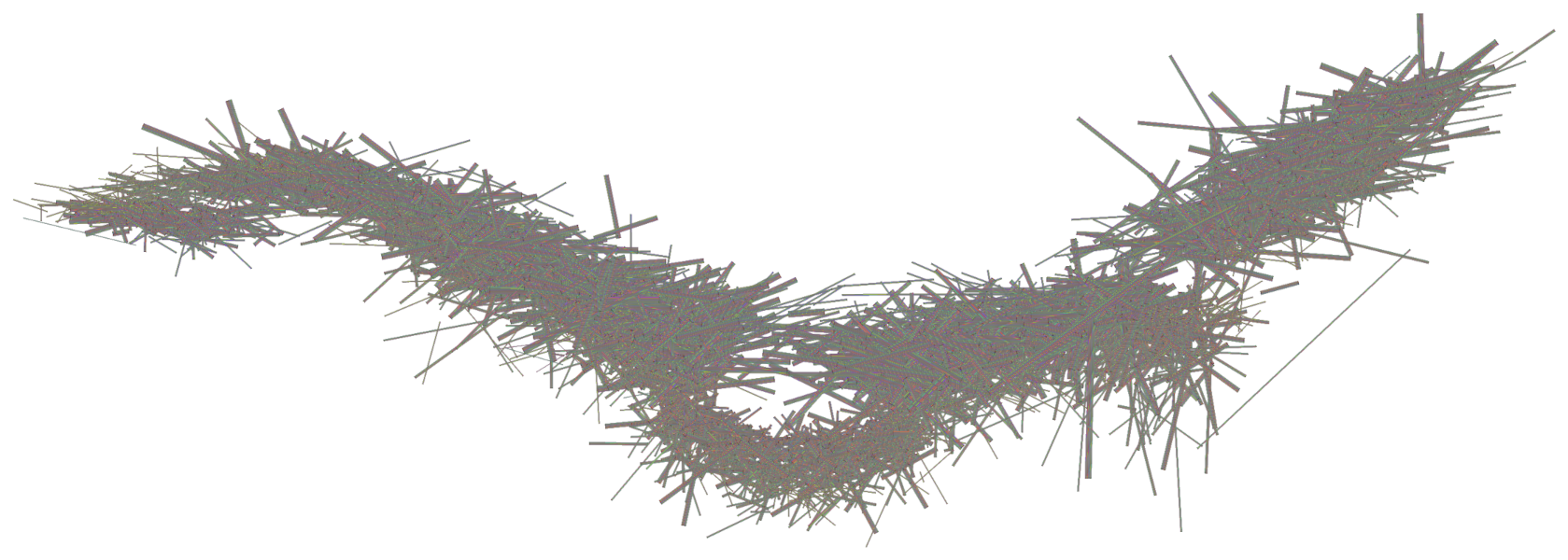}
        \caption{Path stress: $142.2$. }
        \label{fig:DRB1_11}
    \end{subfigure}
    \quad
    \begin{subfigure}[t]{0.45\linewidth}
        \includegraphics[width=\linewidth]{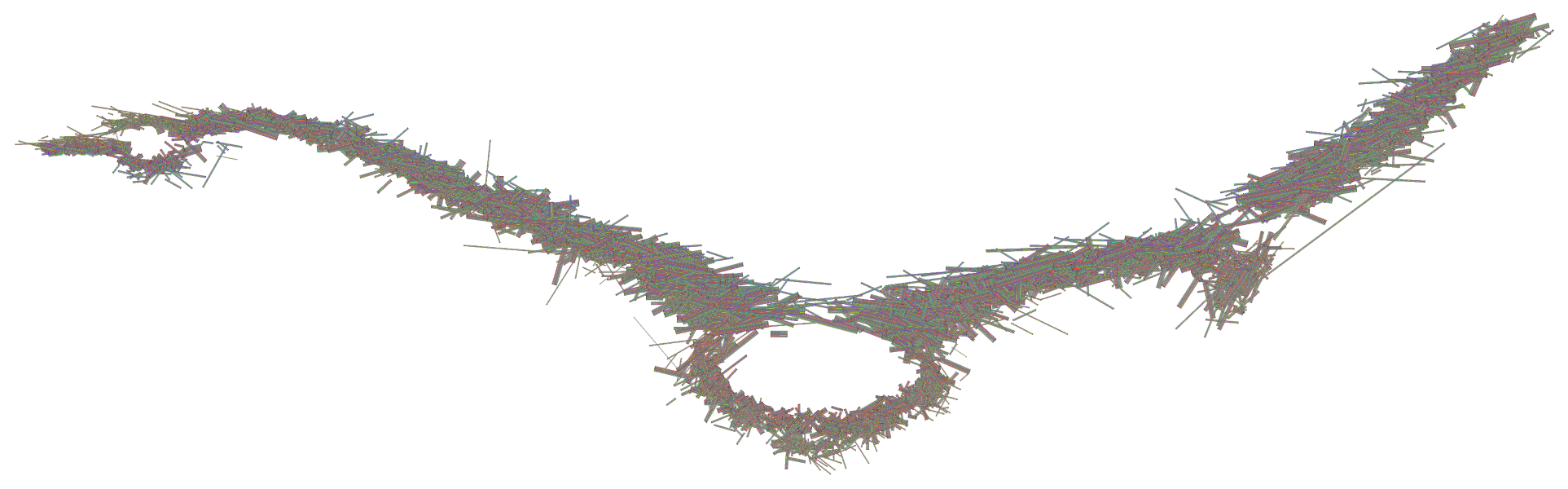}
        \caption{Path stress: $22.4$. }
        \label{fig:DRB1_13}
    \end{subfigure}
    \centering
    \begin{subfigure}[t]{0.45\linewidth}
        \includegraphics[width=\linewidth]{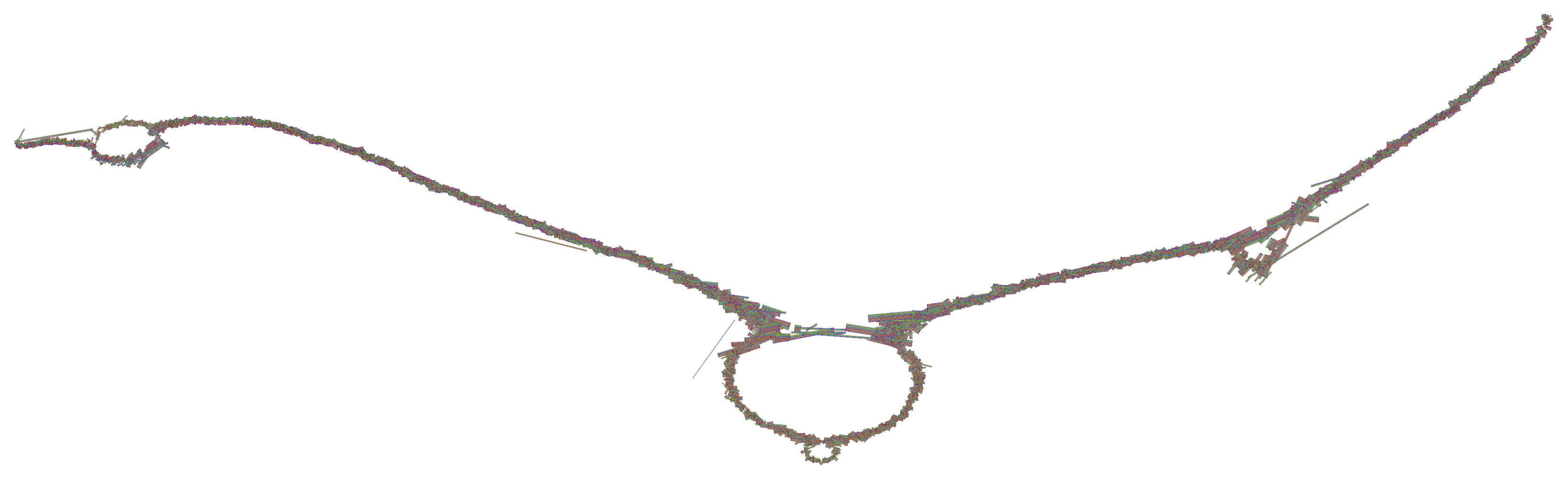}
        \caption{Path stress: $1.3$.  }
        \label{fig:DRB1_15}
    \end{subfigure}
    \quad
    \begin{subfigure}[t]{0.45\linewidth}
        \includegraphics[width=\linewidth]{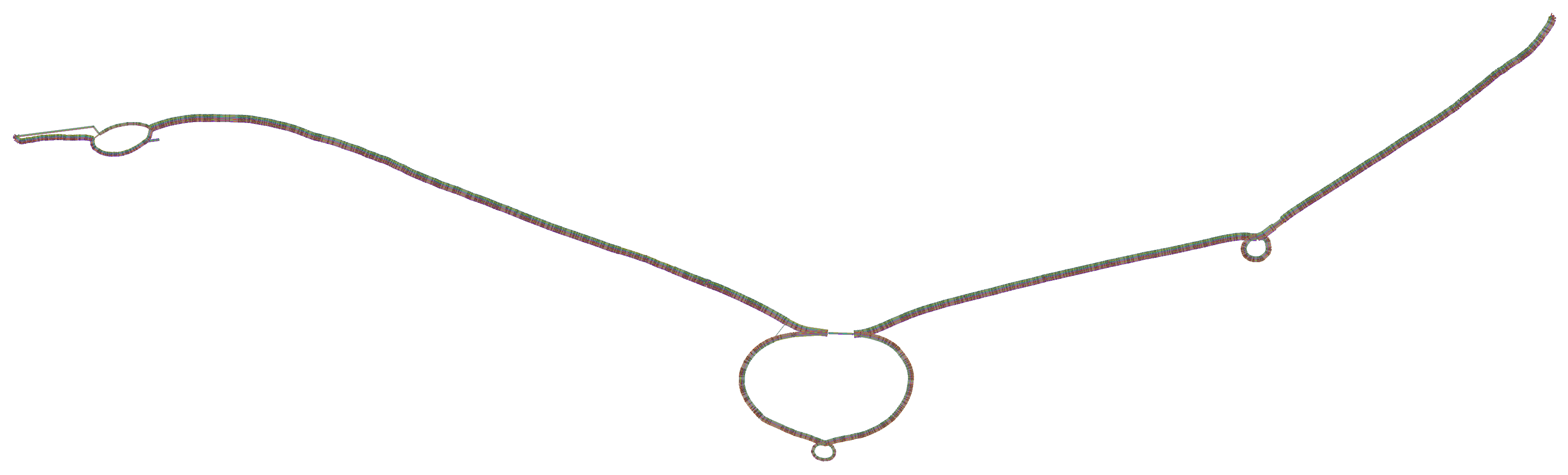}
        \caption{Path stress: $0.07$.  }
        \label{fig:DRB1_29}
    \end{subfigure}    

    \caption{\textbf{Layouts of HLA-DRB1 of different qualities.}}
    \label{fig:metrics-DRB1-different-iter}
\end{figure}

We implement the path stress with a CUDA kernel to speedup the computation by mapping a pair of nodes to each GPU thread, and then aggregating partial results with a reduction tree. 
Fig.~\ref{fig:metrics-DRB1-different-iter} shows how path stress can differentiate between pangenome graph layouts of varying qualities. 
The layout with a lower path stress is considered more legible and aesthetically sound, thereby more effectively revealing the structural information of the pangenome graph. 

\subsection{A Scalable Metric: Sampled Path Stress}
\label{subsec:metrics-sampled-path-stress}

Although path stress can effectively present layout quality, it has a quadratic computational complexity in terms of nodes. 
This poses a significant challenge on scalability even with the compute power of GPUs. As shown in Table~\ref{tab:metrics-runtime-path-stress}, it would require 194 GPU hours with an NVIDIA RTX A6000 GPU to compute the path stress of a human Chr.1 pangenome graph layout, which is impractical. 
Therefore, there is a need for a metric scalable to chromosomal pangenome graphs. 

\begin{table}[!htbp]
    \centering
    \caption{\textbf{Run time of metrics computation.}}
    \label{tab:metrics-runtime-path-stress}
    \resizebox{\columnwidth}{!}{
    \begin{tabular}{l|r|rr}
        \toprule
        Pangenome & \# of Nodes  &  RT of Path Stress &  RT of Sampled Path Stress  \\
        \midrule
        HLA-DRB1 & $5.0\times10^3$ &   1.6 sec	& 0.3 sec	 \\
        MHC      & $2.3\times10^5$ &   53.0 min	& 6.5 sec   \\
        Chr.1    & $1.1\times10^7$ &   (Est.) 194.0 hour	& 5.5 min	 	\\
        \bottomrule
    \end{tabular}
    }
\end{table}

We propose a more scalable metric, \textit{sampled path stress}, which estimates overall path stress by randomly sampling a total of $n$ pairs of visualization nodes $(\mathbf{v_i}, \mathbf{v_j})$, whose corresponding nodes are on the same path.
Equation~\ref{eq:sampled_path_stress} defines sampled path stress, where $\mathcal{S}$ stands for the set of sampled nodes.
By default, we sample $n = 100|p|$ node pairs in each path, where $|p|$ is the number of nodes in path $p$; 
each node is expected to be sampled 100 times within its path. 
\begin{equation}\label{eq:sampled_path_stress}
    \resizebox{0.9\linewidth}{!}{$\displaystyle
        sampled\_path\_stress = \frac{
            \sum_{p \in P} \sum_{(\mathbf{v_i}, \mathbf{v_j)} \in \mathcal{S}} 
            stress_{ij}(\mathbf{v_i}, \mathbf{v_j})
        }{
            n
        }
    $}
\end{equation}

Sampled path stress, which is the mean of the sample (noted as $\mu$), would converge to a normal distribution based on the central limit theorem~\cite{le1986central,kwak2017central}. 
Therefore, we also compute the 95\% confidence interval to validate the sampling coverage. 
This is computed by another pass through the sampled stress terms to get the standard deviation $\sigma$, and derived from $CI_{95\%} = \mu \pm 1.96 {\sigma}/{\sqrt{n}}$.

Applying sampling makes the metric computation linear in complexity, allowing the metric computation for chromosomal pangenome graphs with millions of nodes to be done in minutes, as shown in Table~\ref{tab:metrics-runtime-path-stress}.

\begin{figure}[!htbp]
    \centering
    \includegraphics[width=0.6\linewidth]{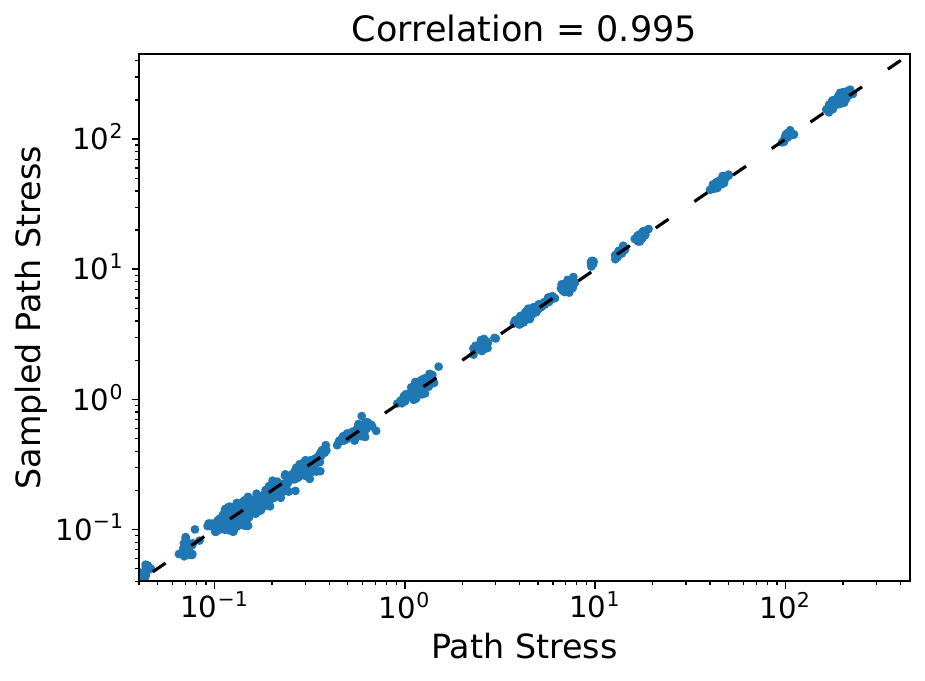}
    \caption{\textbf{Linear correlation} ---  sampled path stress closely approximates the entire path stress. }
    \label{fig:correlation_sampled_stress}
\end{figure}

To check the correctness of sampled path stress, we compare it against path stress with 1824 small-sized pangenome graph layouts, where path stress computation is feasible. Fig.~\ref{fig:correlation_sampled_stress} demonstrates that sampled path stress closely approximates path stress with a correlation of 0.995. 
We also verify that sampled path stress remains consistent with different random seeds for a given layout.

Thus, we adopt sampled path stress as the scalable quantitative metric for evaluating layout quality.

\section{Evaluation}
\label{sec:evaluation}


\subsection{Experiment Setup}
\label{subsec:experiment-setup}

We utilize a 32-core Intel Xeon Gold 6246R CPU@3.4GHz, an NVIDIA RTX A6000 GPU, and an NVIDIA A100 GPU for hardware setup, with GCC 10.2.1 for compilation. 

For overall performance analysis, our GPU design is tested on both an NVIDIA RTX A6000 with CUDA 11.7 and an NVIDIA A100 with CUDA 12.2. 
The ablation study is conducted only on the NVIDIA RTX A6000, utilizing NVIDIA Nsight Compute\cite{nvidia-nsight-compute} 2022.2 and Linux Perf\cite{perf} as profiling tools. 
We perform another case study to explore the performance-quality trade-off with the RTX A6000. 
The multi-threaded CPU baseline is \textit{odgi-layout}~\cite{pangenome-odgi-layout-bioinformatics}.

We use the human pangenome reference dataset released by the HPRC~\cite{draft_human_pangenome_reference}, composed of 24 chromosomal pangenome graphs, from Chr.1 to Chr.22, Chr.X, and Chr.Y. 
As detailed in Table~\ref{tab:dataset}, these graphs contain millions of nodes and are characterized by their notably low node degree and density.

\begin{table}[!htbp]
    \centering
    \caption{\textbf{Properties of the human pangenome graphs.}}
    \label{tab:dataset}
    \resizebox{\columnwidth}{!}{
    \begin{tabular}{l|rrrrrr}
        \toprule
         & \multicolumn{1}{c}{\# Nuc.} & \multicolumn{1}{c}{\# Nodes} & \multicolumn{1}{c}{\# Edges} & \multicolumn{1}{c}{\# Paths}   & \multicolumn{1}{c}{$\overline{deg}$} & \multicolumn{1}{c}{Density} \\
        \midrule
        Min       & $8.8\times10^7$ &	$3.2\times10^5$		& $307$	& $4.4\times10^4$	& $1.4$	& $1.3\times10^{-7}$   \\
        Max       & $1.1\times10^9$	&   $1.1\times10^7$		& $3,029$	& $5.0\times10^5$	& $1.4$	& $4.4\times10^{-6}$   \\
        Mean      & $3.0\times10^8$ &   $4.0\times10^6$		& $1,295$	& $2.3\times10^5$	& $1.4$	& $3.5\times10^{-7}$   \\
        \bottomrule
    \end{tabular}
    }
\end{table}

\subsection{Overall Performance}
\label{subsec:overall-perf}

\begin{table*}[!htbp]
    \centering
    \caption{\textbf {Run time and speedup} --- the run time format is in h:mm:ss.}
    \label{tab:overall_perf}
    \resizebox{\textwidth}{!}{
        \begin{tabular}{lccccc|lccccc|lccccc}
            \toprule
            \multicolumn{1}{c}{Pan.}  & CPU & A6000 & Speedup & A100 & Speedup
                & \multicolumn{1}{c}{Pan.} & CPU & A6000 & Speedup & A100 & Speedup
                    & \multicolumn{1}{c}{Pan.} & CPU & A6000 & Speedup & A100 & Speedup \\
            \midrule
            Chr.1  & 2:32:38  &	0:04:59  & 30.6x  & 0:02:42  & 56.5x  &  Chr.9   & 1:16:49  & 0:02:53  & 26.6x  & 0:00:55  & 83.8x & Chr.17  & 1:03:45	& 0:02:01  & 31.7x  & 0:01:07  & 57.1x  \\
            Chr.2  & 1:17:03  &	0:03:33  & 21.7x  & 0:01:01  & 75.8x  &  Chr.10  & 0:48:34  & 0:02:22  & 20.6x  & 0:00:44  & 66.2x & Chr.18  & 0:50:29	& 0:01:50  & 27.6x  & 0:01:08  & 44.6x  \\
            Chr.3  & 1:28:41  & 0:03:27  & 25.7x  & 0:01:31  & 58.5x  &  Chr.11  & 0:56:25  & 0:02:07  & 26.7x  & 0:00:37  & 91.5x & Chr.19  & 0:40:23	& 0:01:29  & 27.3x  & 0:00:27  & 89.8x  \\
            Chr.4  & 1:47:32  &	0:03:40  & 29.3x  & 0:02:06  & 51.2x  &  Chr.12  & 0:44:05  & 0:02:07  & 20.9x  & 0:00:49  & 54.0x & Chr.20  & 0:51:34	& 0:01:30  & 34.3x  & 0:01:01  & 50.7x  \\
            Chr.5  & 1:41:09  &	0:03:19  & 30.5x  & 0:01:07  & 90.6x  &  Chr.13  & 1:03:32  & 0:02:22  & 26.8x  & 0:00:53  & 71.9x & Chr.21  & 0:44:18	& 0:01:26  & 30.9x  & 0:00:38  & 69.9x  \\
            Chr.6  & 1:13:55  &	0:02:49  & 26.3x  & 0:01:27  & 51.0x  &  Chr.14  & 0:51:21  & 0:02:04  & 24.9x  & 0:00:46  & 67.0x & Chr.22  & 0:39:59	& 0:01:37  & 24.8x  & 0:00:30  & 80.0x  \\
            Chr.7  & 1:16:46  &	0:03:00  & 25.6x  & 0:01:34  & 49.0x  &  Chr.15  & 1:11:33  & 0:02:52  & 25.0x  & 0:01:16  & 56.5x & Chr.X   & 1:04:06	& 0:01:49  & 35.4x  & 0:00:49  & 78.4x  \\
            Chr.8  & 1:17:27  &	0:02:57  & 26.3x  & 0:01:41  & 46.0x  &  Chr.16  & 2:19:47  & 0:04:56  & 28.3x  & 0:12:58  & 10.8x & Chr.Y   & 0:01:55	& 0:00:03  & 36.9x  & 0:00:04  & 28.7x  \\
            \midrule
            \multicolumn{3}{l}{\textbf{Geometric Mean}} & \textbf{27.7x} &   &   \textbf{57.3x} \\
            \bottomrule
        \end{tabular}
    }
    \label{tab:eval-overall-perf}
\end{table*}

\begin{table*}[!htbp]
    \centering
    \caption{\textbf{Layout quality comparison with sampled path stress (SPS)} --- SPS ratio is computed by $\text{GPU}_{\text{SPS}} / \text{CPU}_{\text{SPS}}$.}
    \label{tab:overall_quality}
    \resizebox{\textwidth}{!}{
        \begin{tabular}{lccccc|lccccc|lccccc}
            \toprule
            
            \multicolumn{1}{c}{Pan.}  & CPU $\text{CI}_{95\%}$ & A6000 $\text{CI}_{95\%}$ & SPS ratio & A100 $\text{CI}_{95\%}$ & SPS ratio
                & \multicolumn{1}{c}{Pan.} & CPU $\text{CI}_{95\%}$ & A6000 $\text{CI}_{95\%}$ & SPS ratio & A100 $\text{CI}_{95\%}$ & SPS ratio
                    & \multicolumn{1}{c}{Pan.} & CPU $\text{CI}_{95\%}$ & A6000 $\text{CI}_{95\%}$ & SPS ratio & A100 $\text{CI}_{95\%}$ & SPS ratio \\
                        
            \midrule
            Chr.1  & [0.77, 1.72] &	[0.86, 1.28] &	0.86  & [0.88, 1.28]  & 0.87  & Chr.9   & [0.58, 2.93]   & [0.73, 1.34]  &	0.59 & [-0.15, 3.05] & 0.83  & Chr.17  & [0.45, 0.45]  & [0.60, 0.61]  & 1.34  & [0.58, 0.58]  & 1.29  \\
            Chr.2  & [0.31, 0.76] &	[0.21, 0.29] &	0.47  & [0.35, 0.56]  & 0.85  & Chr.10  & [0.13, 0.17]   & [0.14, 0.19]  &	1.13 & [0.13, 0.17]  & 1.04  & Chr.18  & [0.49, 0.61]  & [0.53, 0.63]  & 1.05  & [0.54, 0.57]  & 1.00  \\
            Chr.3  & [0.26, 0.28] & [0.29, 0.31] &	1.12  & [0.29, 0.30]  & 1.09  & Chr.11  & [0.12, 0.32]   & [0.14, 0.18]  &	0.72 & [0.14, 0.19]  & 0.75  & Chr.19  & [0.17, 0.19]  & [0.22, 0.26]  & 1.30  & [0.19, 0.21]  & 1.11  \\
            Chr.4  & [0.28, 0.31] &	[0.28, 0.30] &	1.00  & [0.28, 0.30]  & 1.00  & Chr.12  & [0.12, 0.17]   & [0.13, 0.14]  &	0.96 & [0.13, 0.16]  & 0.99  & Chr.20  & [0.19, 0.97]  & [0.38, 0.41]  & 0.68  & [0.38, 0.41]  & 0.67  \\
            Chr.5  & [0.18, 0.20] &	[0.22, 0.27] &	1.26  & [0.20, 0.23]  & 1.13  & Chr.13  & [0.38, 0.39]	 & [0.46, 0.51]  &	1.26 & [0.41, 0.48]  & 1.16  & Chr.21  & [0.36, 0.47]  & [0.41, 0.54]  & 1.15  & [0.36, 0.72]  & 1.31  \\
            Chr.6  & [0.30, 0.31] &	[0.32, 0.33] &	1.05  & [0.31, 0.32]  & 1.03  & Chr.14  & [0.19, 0.33]	 & [0.20, 0.88]  &	2.11 & [0.15, 0.61]  & 1.48  & Chr.22  & [0.37, 0.42]  & [0.23, 1.14]  & 1.73  & [0.43, 0.50]  & 1.17  \\
            Chr.7  & [0.28, 0.29] &	[0.29, 0.30] &	1.04  & [0.29, 0.33]  & 1.08  & Chr.15  & [0.85, 1.60]	 & [1.20, 1.85]  &	1.24 & [1.13, 1.67]  & 1.14  & Chr.X   & [0.49, 0.51]  & [0.58, 0.61]  & 1.19  & [0.58, 0.60]  & 1.17  \\
            Chr.8  & [0.27, 0.27] &	[0.27, 0.28] &	1.02  & [0.27, 0.28]  & 1.02  & Chr.16  & [0.67, 0.68]	 & [0.69, 0.71]  &	1.03 & [0.71, 0.72]  & 1.05  & Chr.Y   & [0.58, 0.89]  & [-0.26, 3.66] & 2.31  & [0.63, 0.75]  & 0.94  \\
            \midrule
            \multicolumn{3}{l}{\textbf{Geometric Mean}} & \textbf{1.08} &  & \textbf{1.03}  \\
            \bottomrule
        \end{tabular}
    }
    \label{tab:eval-overall-quality}
\end{table*}

Table~\ref{tab:eval-overall-perf} shows the overall performance for all human chromosomes. 
Our optimized GPU design achieves a 57.3$\times$ speedup on A100, and a 27.7$\times$ speedup on RTX A6000 over the 32-thread CPU baseline \textit{odgi-layout}. 
This reduces the average computation time from 1-2 hours down to just a few minutes. 

\begin{figure}[!htbp]
    \centering
    \begin{subfigure}[t]{\linewidth}
        \includegraphics[width=\linewidth]{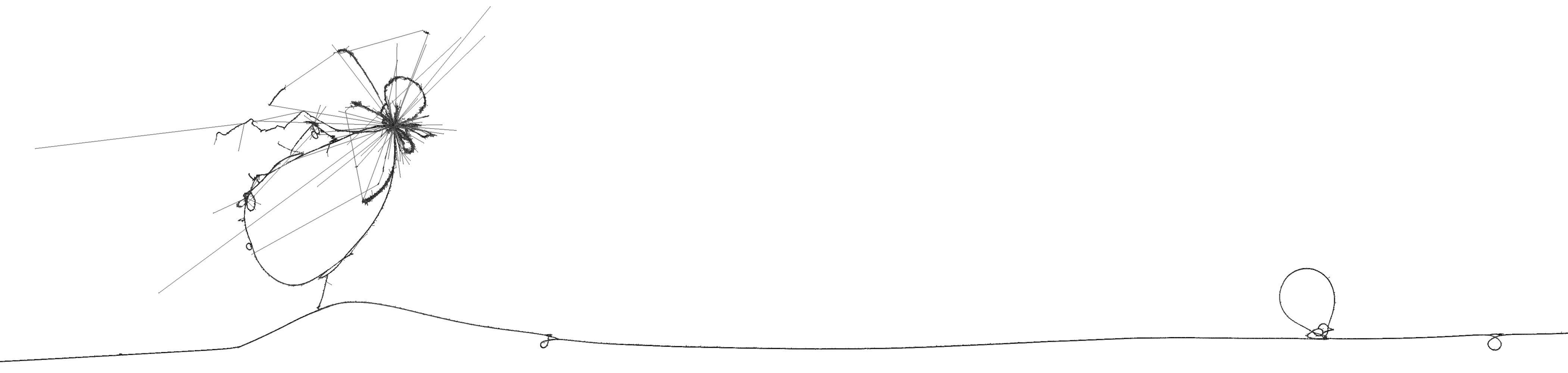}
        \caption{CPU-generated layout.}
        \label{fig:chr7_cpu}
    \end{subfigure}
    \centering
    \begin{subfigure}[t]{\linewidth}
        \includegraphics[width=\linewidth]{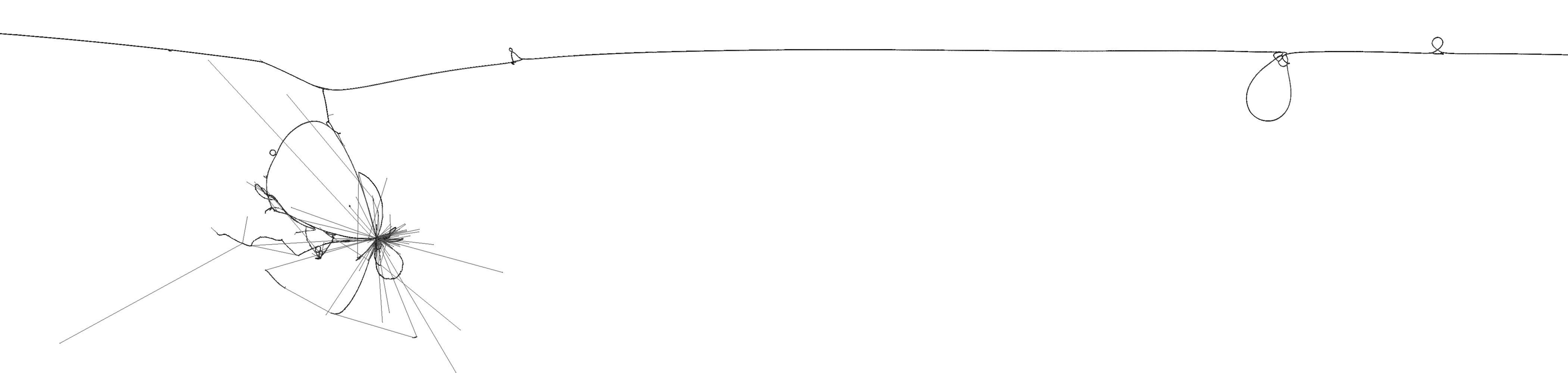}
        \caption{GPU-generated layout.}
        \label{fig:chr7_gpu}
    \end{subfigure}    

    \caption{\textbf{CPU and GPU-generated layouts of Chr.7} --- only the central, most complex parts are displayed, as the entire chromosome is too long. }
    \label{fig:visual-check-chr7}
\end{figure}

We evaluate the layout quality of GPU-generated layouts using both quantitative and qualitative methods. 
Quantitatively, we measure sampled path stress, as shown in Table~\ref{tab:eval-overall-quality}, where the average ratio of sampled path stress between GPU and CPU layouts is close to 1, indicating no quality loss in the GPU-generated layouts. 
Qualitatively, visual inspection confirms that GPU-generated layouts do not have noticeable differences compared to CPU-generated layouts, as demonstrated in Fig.~\ref{fig:visual-check-chr7} for Chr.7. 
Additionally, we conduct 15 runs for each pangenome and confirm the consistency and repeatability of these layouts. 

The 27.7$\times$ speedup achieved by our optimized GPU design on the same NVIDIA RTX A6000 GPU is significantly higher than the 6.8$\times$ improvement achieved by our initial PyTorch implementation, as discussed in Sec.~\ref{subsec:pytorch-design}.
This demonstrates that the custom optimizations in our GPU design effectively exploit the GPU's computing power.

Our design can be seamlessly integrated into the ODGI~\cite{ODGI} framework to facilitate the adoption of our GPU implementation. 
To enable it, a user can simply add the \texttt{--gpu} argument, making the solution effortlessly accessible.

\begin{figure}[!htbp]
    \centering
    \begin{subfigure}[b]{0.49\linewidth}
        \centering
        \includegraphics[width=\linewidth]{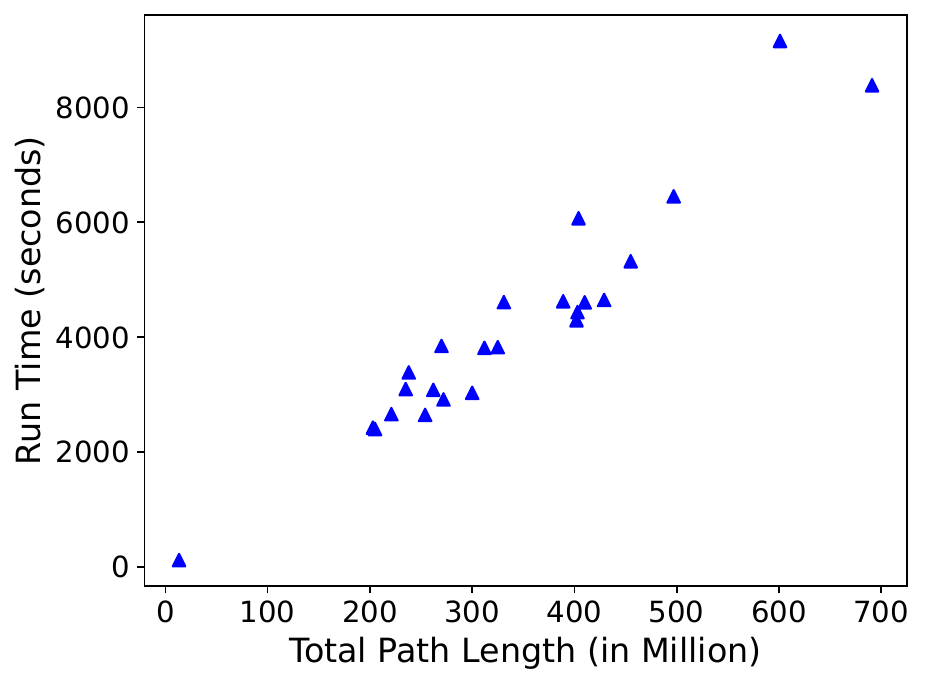}
        \caption{CPU.}
    \end{subfigure}
    \centering
    \begin{subfigure}[b]{0.49\linewidth}
        \centering
        \includegraphics[width=\linewidth]{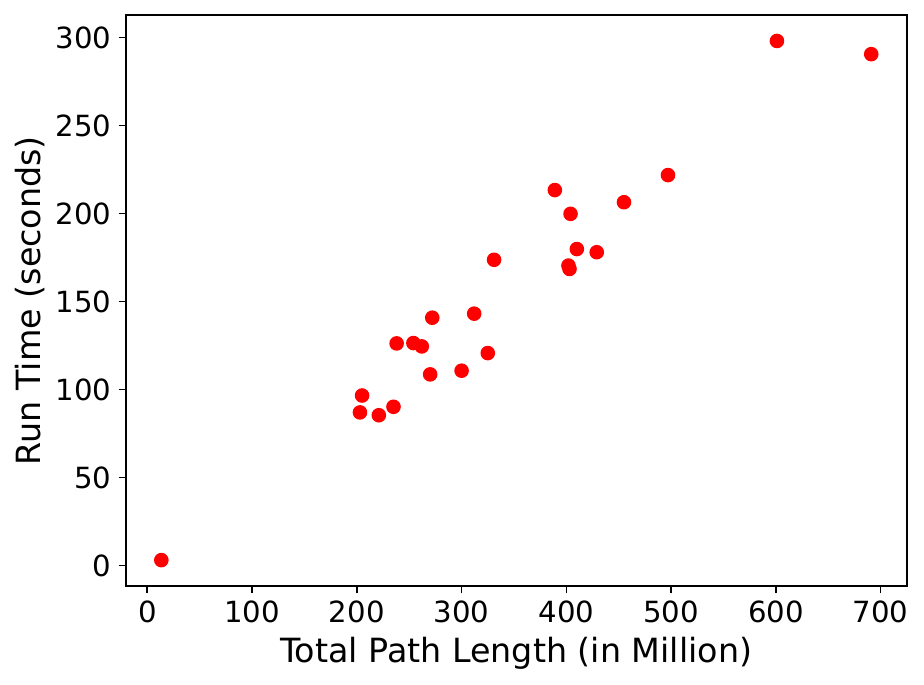}
        \caption{RTX A6000 GPU.}
    \end{subfigure}

    \caption{\textbf{Scalability study on the size of pangenomes.}}
    \label{fig:runtime_scalability_study_x}
\end{figure}

We also perform a scalability study on human pangenome graphs. 
Fig.~\ref{fig:runtime_scalability_study_x} demonstrates linear scaling in both CPU and GPU implementations. 
This result aligns with the expectation, as the number of updates is proportional to the total path length, which is the sum of the nodes in each path.

\subsection{Ablation Study}
\label{subsec:ablation-study}
\begin{figure}[!htbp]
    \centering
    \includegraphics[width=\linewidth]{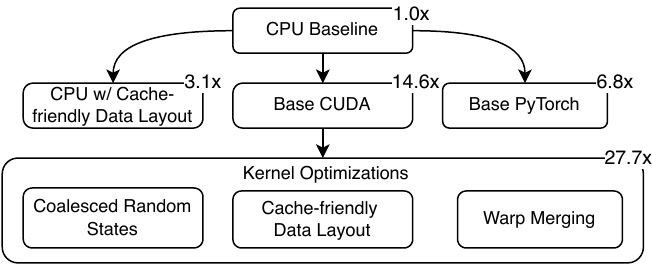}
    \caption{\textbf{Speedup through successive optimizations} -- the baseline is \textit{odgi-layout} on CPU. }
    \label{fig:eval-ablation-overview}
\end{figure}

Fig.~\ref{fig:eval-ablation-overview} shows the incremental performance gains achieved with each optimization. 
Our approach begins with a base CUDA kernel to exploit the data-level parallelism. 
Building on this, we develop an optimized CUDA kernel by introducing three kernel optimization methods.

In the following sections, we evaluate the effects of the individual kernel optimizations.
We apply the methods to the base CUDA kernel individually and evaluate its effects, demonstrating improvements in both run time and key performance metrics.
Since we find similar improvements in all chromosomes, we show here only the results for Chr.1.

\subsubsection{Cache-friendly Data Layout (CDL)}
\label{subsubsec:ablation-cdl}

Since CDL is effective across both CPU and GPU, we apply it on both the base CUDA kernel and the CPU baseline. 
As shown in Table~\ref{tab:ablation-cache-friendly-ds}, 
the improved spatial locality with CDL significantly reduces Last Level Cache (LLC) loads and LLC misses on CPU, and reduces DRAM access on GPU. 

\begin{table}[h]
    \centering
    \caption{\textbf{Effects of cache-friendly data layout}.}
        \begin{tabular}{l|l|rrr}
            \toprule
            \multicolumn{2}{l|}{Method}          & w/o CDL   & w/ CDL    & Improv.    \\
            \midrule
            \multirow{3}{*}{CPU}   & LLC-loads (\#)        & $3.0\times10^{12}$  & $9.4\times10^{11}$   &  $3.2\times$     \\
                                   &  LLC-load-misses (\#) & $2.7\times10^{12}$  & $8.1\times10^{11}$   &  $3.3\times$      \\
                                   &  CPU Run Time (s)     & $9,158.4$            & $2,935.2$             &  $3.1\times$ \\
            \midrule
            \multirow{2}{*}{GPU} & DRAM access (GB)     & $5,191.9$    & $3,974.4$   &  $1.3\times$    \\
                                 & GPU Run Time (s)     & $569.4 $    & $393.1 $   &  $1.4\times$     \\
            \bottomrule
        \end{tabular}
    \label{tab:ablation-cache-friendly-ds}
\end{table}

\subsubsection{Coalesced Random States (CRS)}
\label{subsubsec:ablation-crs}

Table~\ref{tab:ablation-coalesced-random-state} reports the effects of CRS. The ``L1 sectors per request'' metric reflects the level of memory request coalescence within a warp. Here, a request denotes a single instruction requesting a memory operation, and a sector represents an aligned 32B chunk of memory. Each request may access one or more sectors. Hence, fewer sectors per request indicate improved coalescence of memory requests. The CRS method notably decreases the L1 sectors per request, thereby reducing memory accesses to L1, L2 caches, and DRAM.

\begin{table}[!htbp]
    \centering
    \caption{\textbf{Effects of coalesced random states}.}

        \begin{tabular}{l|rrr}
            \toprule
            Method                & w/o CRS   & w/ CRS    & Improv.    \\
            \midrule
            L1 Sectors / Req (\#) & $ 26.8 $    & $9.9   $    &  $2.7\times$     \\
            L1 Cache Access (GB)  & $8,686.7$    & $4,787.7$    &  $1.8\times$      \\
            L2 Cache Access (GB)  & $7,498.9$    & $4,339.3$    &  $1.7\times$      \\
            DRAM Access (GB)      & $5,191.9$    & $4,077.8$    &  $1.3\times$      \\
            GPU Run Time (s)      & $569.4 $    & $471.7 $    &  $1.2\times$     \\
            \bottomrule
        \end{tabular}

    \label{tab:ablation-coalesced-random-state}
\end{table}

\subsubsection{Warp Merging (WM)}
\label{subsubsec:ablation-wm}

As seen in Table~\ref{tab:ablation-warp-merging}, using WM significantly reduces the number of instructions executed. 
The average number of active threads is increased to 27.9, which is close to the full complement of 32 threads within a warp, indicating a considerable reduction in warp divergence.

\begin{table}[!htbp]
    \centering
    \caption{\textbf{Effects of warp merging}.}
    \resizebox{\columnwidth}{!}{
        \begin{tabular}{l|rrr}
            \toprule
            Method                & w/o WM   & w/ WM    & Improv.    \\
            \midrule
            Executed Instructions (\# in billions)  & $131.3$    & $90.1  $    &  $1.5\times$    \\
            Avg. Active Threads Per Warp (\#)       & $20.5 $     &$ 27.9 $     & $ 1.4\times$      \\  
            GPU Run Time (s)                        & $569.4$     &$ 527.4$     & $ 1.1\times$     \\
            \bottomrule
        \end{tabular}
    }

    \label{tab:ablation-warp-merging}
\end{table}

\subsection{A Case Study: Explore the Performance-Quality Trade-off with Sampled Path Stress}
\label{subsec:warp-level-data-reuse}

Sampled path stress proposed in Sec.~\ref{subsec:metrics-sampled-path-stress} allows us to evaluate a chromosomal layout quantitatively in minutes, enabling the exploration of the effects of algorithmic changes on the layout quality of large-scale pangenomes.

While randomness is crucial for achieving high-quality layouts, it limits data reuse, which adversely affects performance. 
We conduct a case study by applying {warp-level data reuse} on top of the optimized GPU design to explore the performance-quality trade-off with sampled path stress. 

\subsubsection{Methods}
\label{subsubsec:case-study-methods}

We aim to increase data reuse with minimal randomness degradation. 
This is accomplished by shuffling node data within the same warp using CUDA  warp-level primitives, enabling direct data sharing between thread registers within the same warp without using shared or global memory. 

The data reuse scheme consists of {data reuse factor (DRF)} and {step reduction factor (SRF)}. 
This modified approach increases the number of updates per step by \textit{DRF}, and reduces the number of steps by \textit{SRF}. 
Each step involves selecting one node pair but performing multiple updates via warp shuffling. 
Through warp shuffling, we reuse the cached data to randomly form a new node pair. 
This method reduces randomness of node selection, potentially affecting the layout quality.

\subsubsection{Results}
\label{subsec:case-study-results}

\begin{figure}[!htbp]

    \centering
    \includegraphics[width=0.5\linewidth]{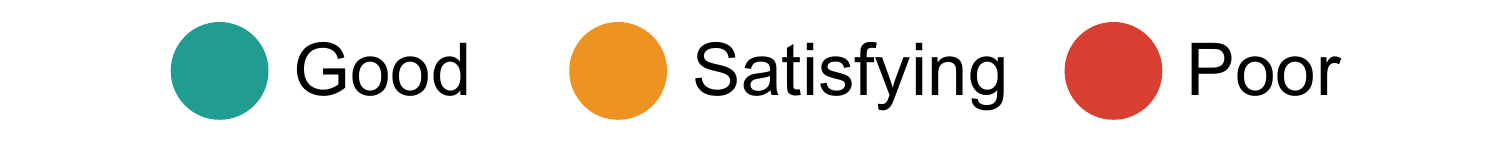}
    \vspace{0.2em}
    \centering
    \begin{subfigure}[b]{0.491\linewidth}
        \centering
        \includegraphics[width=\linewidth]{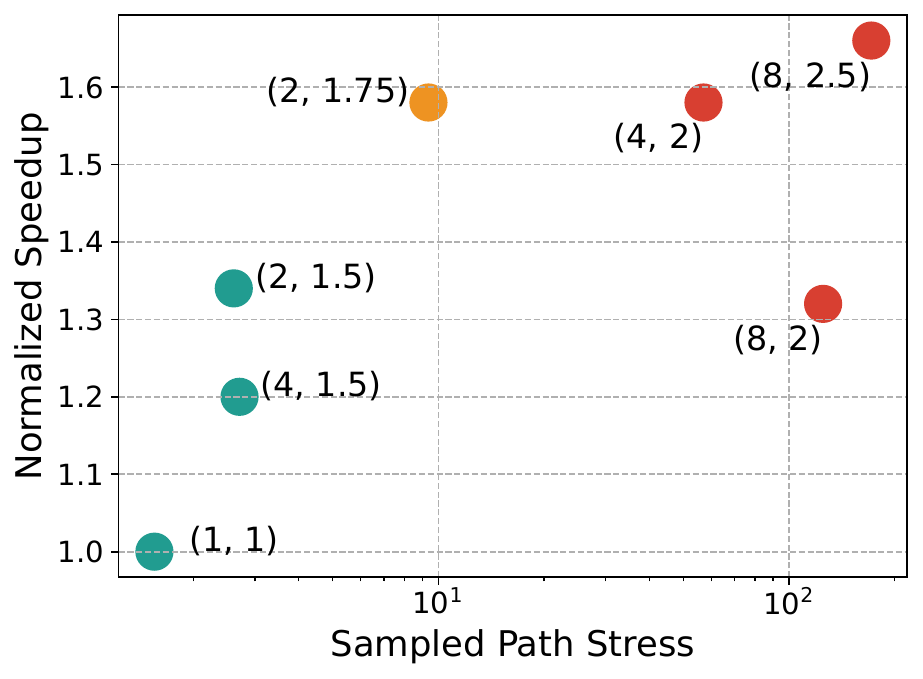}
        \caption{Chr.1.}
    \end{subfigure}
    \centering
    \begin{subfigure}[b]{0.491\linewidth}
        \centering
        \includegraphics[width=\linewidth]{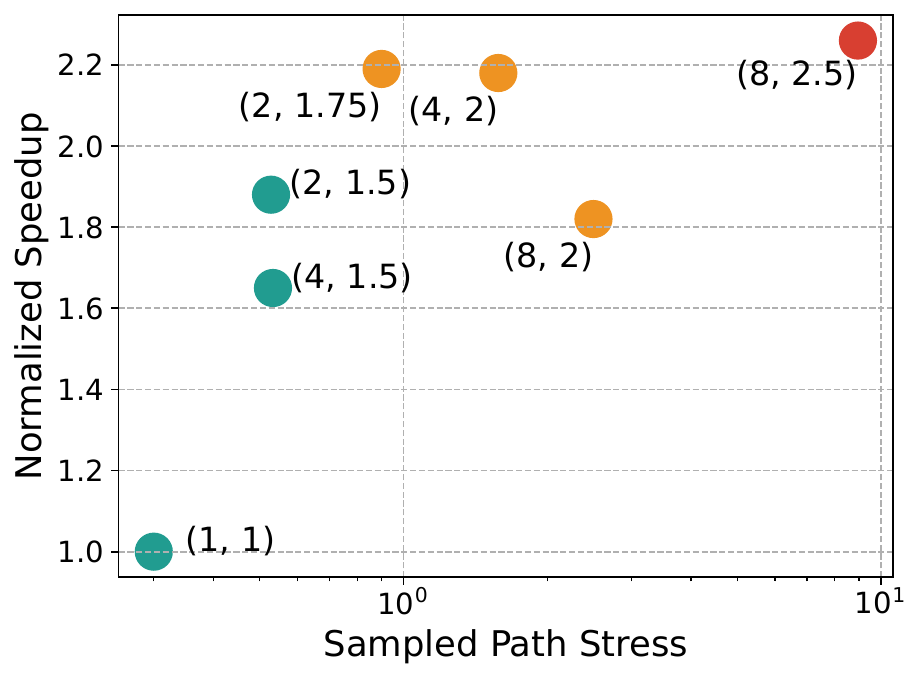}
        \caption{Chr.2.}
    \end{subfigure}
    \caption{\textbf{Design space exploration on data reuse schemes} --- each datapoint represents a $scheme=(DRF, SRF)$. }
    \label{fig:data-reuse-dse}
    
\end{figure}

Fig.~\ref{fig:data-reuse-dse} illustrates the study of the trade-off between performance (represented by normalized speedup) and quality (represented by sampled path stress) in Chr.1 and Chr.2. 
Layouts with stress less than twice that of baseline layouts are considered as ``good'', less than ten times as ``satisfying'', and more than ten times as ``poor''.

An increase in both \textit{DRF} and \textit{SRF} generally leads to an increase in sampled path stress, indicating a loss of layout quality. 
This trend is consistent for most input data.
Overall, schemes with \textit{DRF} of 2 usually produce good or satisfying layouts, while schemes with a \textit{DRF} of 8 tend to produce poor layouts in many cases. 
This is due to reusing data 8 times within a warp's 32 threads, which greatly reduces randomness in node pair selection.

Leveraging scalable sampled path stress allows us to explore the trade-offs between performance and layout quality. 
Tested on the RTX A6000 GPU across all 24 human pangenome graphs, we discover that it is possible to achieve an additional 1.5x speedup over the optimized GPU implementation, while still maintaining good layout quality.

\section{Related Work}
\label{sec:related-work}

Several widely-used graph layout tools include Gephi~\cite{gephi}, NetworkX~\cite{hagberg2008exploring-networkx}, and Graphviz~\cite{ellson2002graphviz}. 
Efforts have been made to accelerate these tools on GPUs, as seen in works like~\cite{brinkmann2017exploiting, fender2022rapids-cugraph}. 
However, the distinct biological meaning of nodes and paths in pangenome graphs limit these tools' suitability.

Numerous pangenome graph layout tools have been developed to better understand the intricate relationships and variations among genomes~\cite{pangenome-graphs}.
AGB~\cite{mikheenko2019AGB} and VG view~\cite{libhandlegraph-garrison2018} employ the rank-based layout algorithm on the Graphviz backend. 
GfaViz~\cite{gonnella2019gfaviz} and BandageNG~\cite{Bandage} adopt the force-directed layout algorithm on the OGDF~\cite{chimani2013open-OGDF} backend. 
SGTK~\cite{kunyavskaya2019sgtk} applies the force-directed layout algorithm on the Cytoscape.js~\cite{franz2016cytoscape} backend. 
Despite their approaches, none have demonstrated scalability to the giga-basepair level, as highlighted in a comprehensive review~\cite{pangenome-graphs} of pangenome graph visualization tools. 
Among the available tools, \textit{odgi-layout}~\cite{pangenome-odgi-layout-bioinformatics} stands to be the only tool capable of scaling to whole-chromosome pangenome graphs containing millions of nodes.  
Despite taking hours, it far surpasses the prior leading tool, BandageNG, which fails to produce a layout within 7 days~\cite{pangenome-odgi-layout-bioinformatics}.

Many efforts have focused on accelerating genomics applications, from GPU-accelerated sequence alignment~\cite{oliveira2016cudalign, goenka2020segalign, zhao2014gblastn, zeni2020logan, muller2022anyseq}, metagenome assembly~\cite{awan2021accelerating} and classification~\cite{kobus2021metacache}, to custom hardware for read assembly~\cite{guo2019gpureadassembly} and read mapping~\cite{cali2022segram}.
Yet, there is a noticeable gap in acceleration work for pangenome graphs, presenting ample opportunity.

\section{Conclusion and Future Work}
\label{sec:conclusion}

We present a fast GPU-based pangenome graph layout solution, achieving a 57.3$\times$ speedup on average over the current state-of-the-art multi-threaded CPU baseline, and 18.5$\times$ speedup over our own optimized version of the CPU solution. 
We leverage the compute power of GPUs with custom optimizations to address the memory-bound and randomness challenges.
Our work enables the layout of the entire human pangenome dataset in just a few minutes, which greatly facilitates pangenomics research. 
For future work, we believe scaling our work to a multi-GPU setup is essential to meet the rapid increase in genome data, and extending to other pangenome analysis applications in the face of increasing data availability.

\section*{Acknowledgment}
For LLM usage disclosure, ChatGPT was utilized to assist and provide suggestions on polishing the text of this paper. 

We are grateful to all the reviewers for their valuable comments. 
This research is supported in part by NSF Awards \#2118709 and \#2118743.

\bibliographystyle{IEEEtran}
\bibliography{references}

@article{garrison2023building,
  title={Building pangenome graphs},
  author = {Erik Garrison and Andrea Guarracino and Simon Heumos and Flavia Villani and Zhigui Bao and Lorenzo Tattini and J{\"o}rg Hagmann and Sebastian Vorbrugg and Santiago Marco-Sola and Christian Kubica and David G. Ashbrook and Kaisa Thorell and Rachel L. Rusholme-Pilcher and Gianni Liti and Emilio Rudbeck and Sven Nahnsen and Zuyu Yang and Mwaniki N. Moses and Franklin L. Nobrega and Yi Wu and Hao Chen and Joep de Ligt and Peter H. Sudmant and Nicole Soranzo and Vincenza Colonna and Robert W. Williams and Pjotr Prins},
  journal={bioRxiv},
  pages={2023--04},
  year={2023},
  publisher={Cold Spring Harbor Laboratory}
}

@article{Guarracino2023,
  doi = {10.1038/s41586-023-05976-y},
  url = {https://doi.org/10.1038/s41586-023-05976-y},
  year = {2023},
  month = may,
  publisher = {Springer Science and Business Media {LLC}},
  volume = {617},
  number = {7960},
  pages = {335--343},
  author = {Andrea Guarracino and Silvia Buonaiuto and Leonardo Gomes de Lima and Tamara Potapova and Arang Rhie and Sergey Koren and Boris Rubinstein and Christian Fischer and Haley J. Abel and Lucinda L. Antonacci-Fulton and Mobin Asri and Gunjan Baid and Carl A. Baker and Anastasiya Belyaeva and Konstantinos Billis and Guillaume Bourque and Andrew Carroll and Mark J. P. Chaisson and Pi-Chuan Chang and Xian H. Chang and Haoyu Cheng and Justin Chu and Sarah Cody and Daniel E. Cook and Robert M. Cook-Deegan and Omar E. Cornejo and Mark Diekhans and Daniel Doerr and Peter Ebert and Jana Ebler and Evan E. Eichler and Jordan M. Eizenga and Susan Fairley and Olivier Fedrigo and Adam L. Felsenfeld and Xiaowen Feng and Paul Flicek and Giulio Formenti and Adam Frankish and Robert S. Fulton and Yan Gao and Shilpa Garg and Nanibaa' A. Garrison and Carlos Garcia Giron and Richard E. Green and Cristian Groza and Leanne Haggerty and Ira Hall and William T. Harvey and Marina Haukness and David Haussler and Simon Heumos and Glenn Hickey and Kendra Hoekzema and Thibaut Hourlier and Kerstin Howe and Miten Jain and Erich D. Jarvis and Hanlee P. Ji and Eimear E. Kenny and Barbara A. Koenig and Alexey Kolesnikov and Jan O. Korbel and Jennifer Kordosky and HoJoon Lee and Alexandra P. Lewis and Heng Li and Wen-Wei Liao and Shuangjia Lu and Tsung-Yu Lu and Julian K. Lucas and Hugo Magalh{\~{a}}es and Santiago Marco-Sola and Pierre Marijon and Charles Markello and Tobias Marschall and Fergal J. Martin and Ann McCartney and Jennifer McDaniel and Karen H. Miga and Matthew W. Mitchell and Jean Monlong and Jacquelyn Mountcastle and Katherine M. Munson and Moses Njagi Mwaniki and Maria Nattestad and Adam M. Novak and Sergey Nurk and Hugh E. Olsen and Nathan D. Olson and Benedict Paten and Trevor Pesout and Alice B. Popejoy and David Porubsky and Pjotr Prins and Daniela Puiu and Mikko Rautiainen and Allison A. Regier and Samuel Sacco and Ashley D. Sanders and Valerie A. Schneider and Baergen I. Schultz and Kishwar Shafin and Jonas A. Sibbesen and Jouni Sir{\'{e}}n and Michael W. Smith and Heidi J. Sofia and Ahmad N. Abou Tayoun and Fran{\c{c}}oise Thibaud-Nissen and Chad Tomlinson and Francesca Floriana Tricomi and Flavia Villani and Mitchell R. Vollger and Justin Wagner and Brian Walenz and Ting Wang and Jonathan M. D. Wood and Aleksey V. Zimin and Justin M. Zook and Jennifer L. Gerton and Adam M. Phillippy and Vincenza Colonna and Erik Garrison},
  title = {Recombination between heterologous human acrocentric chromosomes},
  journal = {Nature}
}

@article{pangenome-odgi-layout-bioinformatics,
    author = {Heumos, Simon and Guarracino, Andrea and Schmelzle, Jan-Niklas M and Li, Jiajie and Zhang, Zhiru and Hagmann, Jörg and Nahnsen, Sven and Prins, Pjotr and Garrison, Erik},
    title = "{Pangenome graph layout by Path-Guided Stochastic Gradient Descent}",
    journal = {Bioinformatics},
    volume = {40},
    number = {7},
    pages = {btae363},
    year = {2024},
    month = {07},
    issn = {1367-4811},
    doi = {10.1093/bioinformatics/btae363},
    url = {https://doi.org/10.1093/bioinformatics/btae363},
    eprint = {https://academic.oup.com/bioinformatics/article-pdf/40/7/btae363/58463786/btae363.pdf},
}

@article{pangenome-graphs,
  title={Pangenome graphs},
  author={Eizenga,  Jordan M and Novak, Adam M and Sibbesen, Jonas A and Heumos, Simon and Ghaffaari, Ali and Hickey, Glenn and Chang, Xian and Seaman, Josiah D and Rounthwaite, Robin and Ebler, Jana and Rautiainen, Mikko and Garg, Shilpa and Paten, Benedict and Marschall, Tobias and Sir\'{e}n, Jouni and Garrison, Erik},
  journal={Annual review of genomics and human genetics},
  volume={21},
  pages={139--162},
  year={2020},
  publisher={Annual Reviews}
}

@article{wang2022human,
  title={The Human Pangenome Project: a global resource to map genomic diversity},
  author={Wang, Ting and Antonacci-Fulton, Lucinda and Howe, Kerstin and Lawson, Heather A and Lucas, Julian K and Phillippy, Adam M and Popejoy, Alice B and Asri, Mobin and Carson, Caryn and Chaisson, Mark JP and and Chang, Xian and Cook-Deegan, Robert and Felsenfeld, Adam L and Fulton, Robert S and Garrison, Erik P and Garrison, Nanibaa' A and Graves-Lindsay, Tina A and Ji, Hanlee and Kenny, Eimear E and Koenig, Barbara A and Li, Daofeng and Marschall, Tobias and McMichael, Joshua F and Novak, Adam M and Purushotham, Deepak and Schneider, Valerie A and Schultz, Baergen I and Smith, Michael W and Sofia, Heidi J and Weissman, Tsachy and Flicek, Paul and Li, Heng and Miga, Karen H and Paten, Benedict and Jarvis, Erich D and Hall, Ira M and Eichler, Evan E and Haussler, David and the Human Pangenome Reference Consortium},
  journal={Nature},
  volume={604},
  number={7906},
  pages={437--446},
  year={2022},
  publisher={Nature Publishing Group UK London}
}

@article{ODGI,
  title={ODGI: understanding pangenome graphs},
  author={Guarracino, Andrea and Heumos, Simon and Nahnsen, Sven and Prins, Pjotr and Garrison, Erik},
  journal={Bioinformatics},
  volume={38},
  number={13},
  pages={3319--3326},
  year={2022},
  publisher={Oxford University Press}
}

@article{variation_graphs,
  title={Efficient dynamic variation graphs},
  author={Eizenga, Jordan M and Novak, Adam M and Kobayashi, Emily and Villani, Flavia and Cisar, Cecilia and Heumos, Simon and Hickey, Glenn and Colonna, Vincenza and Paten, Benedict and Garrison, Erik},
  journal={Bioinformatics},
  volume={36},
  number={21},
  pages={5139--5144},
  year={2020},
  publisher={Oxford University Press}
}

@article{draft_human_pangenome_reference,
  title={A draft human pangenome reference},
  author={Liao, Wen-Wei and Asri, Mobin and Ebler, Jana and Doerr, Daniel and Haukness, Marina and Hickey, Glenn and Lu, Shuangjia and Lucas, Julian K and Monlong, Jean and Abel, Haley J and and Buonaiuto, Silvia and Chang, Xian H and Cheng, Haoyu and Chu, Justin and Colonna, Vincenza and Eizenga, Jordan M and Feng, Xiaowen and Fischer, Christian and Fulton, Robert S and Garg, Shilpa and Groza, Cristian and Guarracino, Andrea and Harvey, William T and Heumos, Simon and Howe, Kerstin and Jain, Miten and Lu, Tsung-Yu and Markello, Charles and Martin, Fergal J and Mitchell, Matthew W and Munson, Katherine M and Mwaniki, Moses Njagi and Novak, Adam M and Olsen, Hugh E and Pesout, Trevor and Porubsky, David and Prins, Pjotr and Sibbesen, Jonas A and Sir{\'e}n, Jouni and Tomlinson, Chad and Villani, Flavia and Vollger, Mitchell R and Antonacci-Fulton, Lucinda L and Baid, Gunjan and Baker, Carl A and Belyaeva, Anastasiya and Billis, Konstantinos and Carroll, Andrew and Chang, Pi-Chuan and Cody, Sarah and Cook, Daniel E and Cook-Deegan, Robert M and Cornejo, Omar E and Diekhans, Mark and Ebert, Peter and Fairley, Susan and Fedrigo, Olivier and Felsenfeld, Adam L and Formenti, Giulio and Frankish, Adam and Gao, Yan and Garrison, Nanibaa' A and Giron, Carlos Garcia and Green, Richard E and Haggerty, Leanne and Hoekzema, Kendra and Hourlier, Thibaut and Ji, Hanlee P and Kenny, Eimear E and Koenig, Barbara A and Kolesnikov, Alexey and Korbel, Jan O and Kordosky, Jennifer and Koren, Sergey and Lee, HoJoon and Lewis, Alexandra P and Magalh{\~a}es, Hugo and Marco-Sola, Santiago and Marijon, Pierre and McCartney, Ann and McDaniel, Jennifer and Mountcastle, Jacquelyn and Nattestad, Maria and Nurk, Sergey and Olson, Nathan D and Popejoy, Alice B and Puiu, Daniela and Rautiainen, Mikko and Regier, Allison A and Rhie, Arang and Sacco, Samuel and Sanders, Ashley D and Schneider, Valerie A and Schultz, Baergen I and Shafin, Kishwar and Smith, Michael W and Sofia, Heidi J and Abou Tayoun, Ahmad N and Thibaud-Nissen, Fran{\c{c}}oise and Tricomi, Francesca Floriana and Wagner, Justin and Walenz, Brian and Wood, Jonathan M D and Zimin, Aleksey V and Bourque, Guillaume and Chaisson, Mark J P and Flicek, Paul and Phillippy, Adam M and Zook, Justin M and Eichler, Evan E and Haussler, David and Wang, Ting and Jarvis, Erich D and Miga, Karen H and Garrison, Erik and Marschall, Tobias and Hall, Ira M and Li, Heng and Paten, Benedict},
  journal={Nature},
  volume={617},
  number={7960},
  pages={312--324},
  year={2023},
  publisher={Nature Publishing Group UK London}
}

@article{eisenstein2023every,
  title={Every base everywhere all at once: Pangenomics comes of age},
  author={Eisenstein, Michael},
  journal={Nature},
  volume={616},
  number={7957},
  pages={618--620},
  year={2023},
  publisher={Nature}
}

@article{ballouz2019time,
  title={Is it time to change the reference genome?},
  author={Ballouz, Sara and Dobin, Alexander and Gillis, Jesse A},
  journal={Genome biology},
  volume={20},
  number={1},
  pages={1--9},
  year={2019},
  publisher={BioMed Central}
}

@article{golicz2020pangenomics,
  title={Pangenomics comes of age: from bacteria to plant and animal applications},
  author={Golicz, Agnieszka A and Bayer, Philipp E and Bhalla, Prem L and Batley, Jacqueline and Edwards, David},
  journal={Trends in Genetics},
  volume={36},
  number={2},
  pages={132--145},
  year={2020},
  publisher={Elsevier}
}

@article{paten2017genome,
  title={Genome graphs and the evolution of genome inference},
  author={Paten, Benedict and Novak, Adam M and Eizenga, Jordan M and Garrison, Erik},
  journal={Genome research},
  volume={27},
  number={5},
  pages={665--676},
  year={2017},
  publisher={Cold Spring Harbor Lab}
}

@article{yang2023neisseriameningitidis,
  title={Pangenome graphs in infectious disease: a comprehensive genetic variation analysis of Neisseria meningitidis leveraging Oxford Nanopore long reads},
  author={Yang, Zuyu and Guarracino, Andrea and Biggs, Patrick J and Black, Michael A and Ismail, Nuzla and Wold, Jana Renee and Merriman, Tony R and Prins, Pjotr and Garrison, Erik and de Ligt, Joep},
  journal={Frontiers in Genetics},
  volume={14},
  year={2023},
  publisher={Frontiers Media SA},
  DOI={10.3389/fgene.2023.1225248}
}

@article{groza2023geneticdiseases,
  title={Pangenome graphs improve the analysis of rare genetic diseases},
  author={Groza, Cristian and Schwendinger-Schreck, Carl and Cheung, Warren A and Farrow, Emily G and Thiffault, Isabelle and Lake, Juniper and Rizzo, William B and Evrony, Gilad and Curran, Tom and Bourque, Guillaume and Pastinen, Tomi},
  journal={medRxiv},
  pages={2023--05},
  year={2023},
  publisher={Cold Spring Harbor Laboratory Press}
}

@article{hubner2022we,
  title={Are we there yet? Driving the road to evolutionary graph-pangenomics},
  author={H{\"u}bner, Sariel},
  journal={Current Opinion in Plant Biology},
  volume={66},
  pages={102195},
  year={2022},
  publisher={Elsevier}
}

@article{libhandlegraph-garrison2018,
  title={Variation graph toolkit improves read mapping by representing genetic variation in the reference},
  author={Garrison, Erik and Sir{\'e}n, Jouni and Novak, Adam M and Hickey, Glenn and Eizenga, Jordan M and Dawson, Eric T and Jones, William and Garg, Shilpa and Markello, Charles and Lin, Michael F and Paten, Benedict and Durbin, Richard},
  journal={Nature biotechnology},
  volume={36},
  number={9},
  pages={875--879},
  year={2018},
  publisher={Nature Publishing Group US New York}
}

@inproceedings{goenka2020segalign,
  title={SegAlign: A scalable GPU-based whole genome aligner},
  author={Goenka, Sneha D and Turakhia, Yatish and Paten, Benedict and Horowitz, Mark},
  booktitle={SC20: International Conference for High Performance Computing, Networking, Storage and Analysis},
  pages={1--13},
  year={2020},
  organization={IEEE}
}

@article{zhao2014gblastn,
  title={G-BLASTN: accelerating nucleotide alignment by graphics processors},
  author={Zhao, Kaiyong and Chu, Xiaowen},
  journal={Bioinformatics},
  volume={30},
  number={10},
  pages={1384--1391},
  year={2014},
  publisher={Oxford University Press}
}

@article{oliveira2016cudalign,
  title={CUDAlign 4.0: Incremental speculative traceback for exact chromosome-wide alignment in GPU clusters},
  author={de Oliveira Sandes, Edans Flavius and Miranda, Guillermo and Martorell, Xavier and Ayguade, Eduard and Teodoro, George and Melo, Alba Cristina Magalhaes},
  journal={IEEE Transactions on Parallel and Distributed Systems},
  volume={27},
  number={10},
  pages={2838--2850},
  year={2016},
  publisher={IEEE}
}

@inproceedings{guo2019gpureadassembly,
  title={Hardware acceleration of long read pairwise overlapping in genome sequencing: A race between fpga and gpu},
  author={Guo, Licheng and Lau, Jason and Ruan, Zhenyuan and Wei, Peng and Cong, Jason},
  booktitle={2019 IEEE 27th Annual International Symposium on Field-Programmable Custom Computing Machines (FCCM)},
  pages={127--135},
  year={2019},
  organization={IEEE}
}

@article{first-human-genome,
  title={The sequence of the human genome},
  author={J. Craig Venter  and Mark D. Adams  and Eugene W. Myers  and Peter W. Li  and Richard J. Mural  and Granger G. Sutton  and Hamilton O. Smith  and Mark Yandell  and Cheryl A. Evans  and Robert A. Holt  and Jeannine D. Gocayne  and Peter Amanatides  and Richard M. Ballew  and Daniel H. Huson  and Jennifer Russo Wortman  and Qing Zhang  and Chinnappa D. Kodira  and Xiangqun H. Zheng  and Lin Chen  and Marian Skupski  and Gangadharan Subramanian  and Paul D. Thomas  and Jinghui Zhang  and George L. Gabor Miklos  and Catherine Nelson  and Samuel Broder  and Andrew G. Clark  and Joe Nadeau  and Victor A. McKusick  and Norton Zinder  and Arnold J. Levine  and Richard J. Roberts  and Mel Simon  and Carolyn Slayman  and Michael Hunkapiller  and Randall Bolanos  and Arthur Delcher  and Ian Dew  and Daniel Fasulo  and Michael Flanigan  and Liliana Florea  and Aaron Halpern  and Sridhar Hannenhalli  and Saul Kravitz  and Samuel Levy  and Clark Mobarry  and Knut Reinert  and Karin Remington  and Jane Abu-Threideh  and Ellen Beasley  and Kendra Biddick  and Vivien Bonazzi  and Rhonda Brandon  and Michele Cargill  and Ishwar Chandramouliswaran  and Rosane Charlab  and Kabir Chaturvedi  and Zuoming Deng  and Valentina Di Francesco  and Patrick Dunn  and Karen Eilbeck  and Carlos Evangelista  and Andrei E. Gabrielian  and Weiniu Gan  and Wangmao Ge  and Fangcheng Gong  and Zhiping Gu  and Ping Guan  and Thomas J. Heiman  and Maureen E. Higgins  and Rui-Ru Ji  and Zhaoxi Ke  and Karen A. Ketchum  and Zhongwu Lai  and Yiding Lei  and Zhenya Li  and Jiayin Li  and Yong Liang  and Xiaoying Lin  and Fu Lu  and Gennady V. Merkulov  and Natalia Milshina  and Helen M. Moore  and Ashwinikumar K Naik  and Vaibhav A. Narayan  and Beena Neelam  and Deborah Nusskern  and Douglas B. Rusch  and Steven Salzberg  and Wei Shao  and Bixiong Shue  and Jingtao Sun  and Zhen Yuan Wang  and Aihui Wang  and Xin Wang  and Jian Wang  and Ming-Hui Wei  and Ron Wides  and Chunlin Xiao  and Chunhua Yan  and Alison Yao  and Jane Ye  and Ming Zhan  and Weiqing Zhang  and Hongyu Zhang  and Qi Zhao  and Liansheng Zheng  and Fei Zhong  and Wenyan Zhong  and Shiaoping C. Zhu  and Shaying Zhao  and Dennis Gilbert  and Suzanna Baumhueter  and Gene Spier  and Christine Carter  and Anibal Cravchik  and Trevor Woodage  and Feroze Ali  and Huijin An  and Aderonke Awe  and Danita Baldwin  and Holly Baden  and Mary Barnstead  and Ian Barrow  and Karen Beeson  and Dana Busam  and Amy Carver  and Angela Center  and Ming Lai Cheng  and Liz Curry  and Steve Danaher  and Lionel Davenport  and Raymond Desilets  and Susanne Dietz  and Kristina Dodson  and Lisa Doup  and Steven Ferriera  and Neha Garg  and Andres Gluecksmann  and Brit Hart  and Jason Haynes  and Charles Haynes  and Cheryl Heiner  and Suzanne Hladun  and Damon Hostin  and Jarrett Houck  and Timothy Howland  and Chinyere Ibegwam  and Jeffery Johnson  and Francis Kalush  and Lesley Kline  and Shashi Koduru  and Amy Love  and Felecia Mann  and David May  and Steven McCawley  and Tina McIntosh  and Ivy McMullen  and Mee Moy  and Linda Moy  and Brian Murphy  and Keith Nelson  and Cynthia Pfannkoch  and Eric Pratts  and Vinita Puri  and Hina Qureshi  and Matthew Reardon  and Robert Rodriguez  and Yu-Hui Rogers  and Deanna Romblad  and Bob Ruhfel  and Richard Scott  and Cynthia Sitter  and Michelle Smallwood  and Erin Stewart  and Renee Strong  and Ellen Suh  and Reginald Thomas  and Ni Ni Tint  and Sukyee Tse  and Claire Vech  and Gary Wang  and Jeremy Wetter  and Sherita Williams  and Monica Williams  and Sandra Windsor  and Emily Winn-Deen  and Keriellen Wolfe  and Jayshree Zaveri  and Karena Zaveri  and Josep F. Abril  and Roderic Guigó  and Michael J. Campbell  and Kimmen V. Sjolander  and Brian Karlak  and Anish Kejariwal  and Huaiyu Mi  and Betty Lazareva  and Thomas Hatton  and Apurva Narechania  and Karen Diemer  and Anushya Muruganujan  and Nan Guo  and Shinji Sato  and Vineet Bafna  and Sorin Istrail  and Ross Lippert  and Russell Schwartz  and Brian Walenz  and Shibu Yooseph  and David Allen  and Anand Basu  and James Baxendale  and Louis Blick  and Marcelo Caminha  and John Carnes-Stine  and Parris Caulk  and Yen-Hui Chiang  and My Coyne  and Carl Dahlke  and Anne Deslattes Mays  and Maria Dombroski  and Michael Donnelly  and Dale Ely  and Shiva Esparham  and Carl Fosler  and Harold Gire  and Stephen Glanowski  and Kenneth Glasser  and Anna Glodek  and Mark Gorokhov  and Ken Graham  and Barry Gropman  and Michael Harris  and Jeremy Heil  and Scott Henderson  and Jeffrey Hoover  and Donald Jennings  and Catherine Jordan  and James Jordan  and John Kasha  and Leonid Kagan  and Cheryl Kraft  and Alexander Levitsky  and Mark Lewis  and Xiangjun Liu  and John Lopez  and Daniel Ma  and William Majoros  and Joe McDaniel  and Sean Murphy  and Matthew Newman  and Trung Nguyen  and Ngoc Nguyen  and Marc Nodell  and Sue Pan  and Jim Peck  and Marshall Peterson  and William Rowe  and Robert Sanders  and John Scott  and Michael Simpson  and Thomas Smith  and Arlan Sprague  and Timothy Stockwell  and Russell Turner  and Eli Venter  and Mei Wang  and Meiyuan Wen  and David Wu  and Mitchell Wu  and Ashley Xia  and Ali Zandieh  and Xiaohong Zhu},
  journal={science},
  volume={291},
  number={5507},
  pages={1304--1351},
  year={2001},
  publisher={American Association for the Advancement of Science}
}

@article{HLA-DRB1-covid19,
  title={The influence of HLA genotype on the severity of COVID-19 infection},
  author={Langton, David J and Bourke, Stephen C and Lie, Benedicte A and Reiff, Gabrielle and Natu, Shonali and Darlay, Rebecca and Burn, John and Echevarria, Carlos},
  journal={Hla},
  volume={98},
  number={1},
  pages={14--22},
  year={2021},
  publisher={Wiley Online Library}
}

@article{shiina2009hla,
  title={The HLA genomic loci map: expression, interaction, diversity and disease},
  author={Shiina, Takashi and Hosomichi, Kazuyoshi and Inoko, Hidetoshi and Kulski, Jerzy K},
  journal={Journal of human genetics},
  volume={54},
  number={1},
  pages={15--39},
  year={2009},
  publisher={Nature Publishing Group}
}

@article{rautiainen2023telomere,
  title={Telomere-to-telomere assembly of diploid chromosomes with Verkko},
  author={Rautiainen, Mikko and Nurk, Sergey and Walenz, Brian P and Logsdon, Glennis A and Porubsky, David and Rhie, Arang and Eichler, Evan E and Phillippy, Adam M and Koren, Sergey},
  journal={Nature Biotechnology},
  pages={1--9},
  year={2023},
  publisher={Nature Publishing Group US New York}
}

@article{cheng2023scalable,
  title={Scalable telomere-to-telomere assembly for diploid and polyploid genomes with double graph},
  author={Cheng, Haoyu and Asri, Mobin and Lucas, Julian and Koren, Sergey and Li, Heng},
  journal={arXiv preprint arXiv:2306.03399},
  year={2023}
}

@inproceedings{awan2021accelerating,
  title={Accelerating large scale de novo metagenome assembly using GPUs},
  author={Awan, Muaaz Gul and Hofmeyr, Steven and Egan, Rob and Ding, Nan and Buluc, Aydin and Deslippe, Jack and Oliker, Leonid and Yelick, Katherine},
  booktitle={Proceedings of the International Conference for High Performance Computing, Networking, Storage and Analysis},
  pages={1--11},
  year={2021}
}

@inproceedings{muller2022anyseq,
  title={AnySeq/GPU: a novel approach for faster sequence alignment on GPUs},
  author={M{\"u}ller, Andr{\'e} and Schmidt, Bertil and Membarth, Richard and Lei{\ss}a, Roland and Hack, Sebastian},
  booktitle={Proceedings of the 36th ACM International Conference on Supercomputing},
  pages={1--11},
  year={2022}
}

@inproceedings{kobus2021metacache,
  title={MetaCache-GPU: ultra-fast metagenomic classification},
  author={Kobus, Robin and M{\"u}ller, Andr{\'e} and J{\"u}nger, Daniel and Hundt, Christian and Schmidt, Bertil},
  booktitle={Proceedings of the 50th International Conference on Parallel Processing},
  pages={1--11},
  year={2021}
}

@article{Bandage,
  title={Bandage: interactive visualization of de novo genome assemblies},
  author={Wick, Ryan R and Schultz, Mark B and Zobel, Justin and Holt, Kathryn E},
  journal={Bioinformatics},
  volume={31},
  number={20},
  pages={3350--3352},
  year={2015},
  publisher={Oxford University Press}
}

@article{gonnella2019gfaviz,
  title={GfaViz: flexible and interactive visualization of GFA sequence graphs},
  author={Gonnella, Giorgio and Niehus, Niklas and Kurtz, Stefan},
  journal={Bioinformatics},
  volume={35},
  number={16},
  pages={2853--2855},
  year={2019},
  publisher={Oxford University Press}
}

@article{mikheenko2019AGB,
  title={Assembly Graph Browser: interactive visualization of assembly graphs},
  author={Mikheenko, Alla and Kolmogorov, Mikhail},
  journal={Bioinformatics},
  volume={35},
  number={18},
  pages={3476--3478},
  year={2019},
  publisher={Oxford University Press}
}

@article{kunyavskaya2019sgtk,
  title={SGTK: a toolkit for visualization and assessment of scaffold graphs},
  author={Kunyavskaya, Olga and Prjibelski, Andrey D},
  journal={Bioinformatics},
  volume={35},
  number={13},
  pages={2303--2305},
  year={2019},
  publisher={Oxford University Press}
}

@article{graphdraw-by-SGD,
  title={Graph drawing by stochastic gradient descent},
  author={Zheng, Jonathan X and Pawar, Samraat and Goodman, Dan FM},
  journal={IEEE transactions on visualization and computer graphics},
  volume={25},
  number={9},
  pages={2738--2748},
  year={2018},
  publisher={IEEE}
}

@article{hogwild,
  title={Hogwild!: A lock-free approach to parallelizing stochastic gradient descent},
  author={Recht, Benjamin and Re, Christopher and Wright, Stephen and Niu, Feng},
  journal={Advances in neural information processing systems},
  volume={24},
  year={2011}
}

@inproceedings{cali2022segram,
  title={SeGraM: A universal hardware accelerator for genomic sequence-to-graph and sequence-to-sequence mapping},
  author={Cali, Damla Senol and Kanellopoulos, Konstantinos and Lindegger, Jo{\"e}l and Bing{\"o}l, Z{\"u}lal and Kalsi, Gurpreet S and Zuo, Ziyi and Firtina, Can and Cavlak, Meryem Banu and Kim, Jeremie and Ghiasi, Nika Mansouri and others},
  booktitle={Proceedings of the 49th Annual International Symposium on Computer Architecture},
  pages={638--655},
  year={2022}
}

@inproceedings{zeni2020logan,
  title={Logan: High-performance gpu-based x-drop long-read alignment},
  author={Zeni, Alberto and Guidi, Giulia and Ellis, Marquita and Ding, Nan and Santambrogio, Marco D and Hofmeyr, Steven and Bulu{\c{c}}, Ayd{\i}n and Oliker, Leonid and Yelick, Katherine},
  booktitle={2020 IEEE International Parallel and Distributed Processing Symposium (IPDPS)},
  pages={462--471},
  year={2020},
  organization={IEEE}
}

@article{franz2016cytoscape,
  title={Cytoscape. js: a graph theory library for visualisation and analysis},
  author={Franz, Max and Lopes, Christian T and Huck, Gerardo and Dong, Yue and Sumer, Onur and Bader, Gary D},
  journal={Bioinformatics},
  volume={32},
  number={2},
  pages={309--311},
  year={2016},
  publisher={Oxford University Press}
}

@article{chimani2013open-OGDF,
  title={The Open Graph Drawing Framework (OGDF).},
  author={Chimani, Markus and Gutwenger, Carsten and J{\"u}nger, Michael and Klau, Gunnar W and Klein, Karsten and Mutzel, Petra},
  journal={Handbook of graph drawing and visualization},
  volume={2011},
  pages={543--569},
  year={2013}
}

@inproceedings{brinkmann2017exploiting,
  title={Exploiting GPUs for fast force-directed visualization of large-scale networks},
  author={Brinkmann, Govert G and Rietveld, Kristian FD and Takes, Frank W},
  booktitle={2017 46th International Conference on Parallel Processing (ICPP)},
  pages={382--391},
  year={2017},
  organization={IEEE}
}

@incollection{fender2022rapids-cugraph,
  title={Rapids cugraph},
  author={Fender, Alex and Rees, Brad and Eaton, Joe},
  booktitle={Massive Graph Analytics},
  pages={483--493},
  year={2022},
  publisher={Chapman and Hall/CRC}
}

@inproceedings{ellson2002graphviz,
  title={Graphviz—open source graph drawing tools},
  author={Ellson, John and Gansner, Emden and Koutsofios, Lefteris and North, Stephen C and Woodhull, Gordon},
  booktitle={Graph Drawing: 9th International Symposium, GD 2001 Vienna, Austria, September 23--26, 2001 Revised Papers 9},
  pages={483--484},
  year={2002},
  organization={Springer}
}

@techreport{hagberg2008exploring-networkx,
  title={Exploring network structure, dynamics, and function using NetworkX},
  author={Hagberg, Aric and Swart, Pieter and S Chult, Daniel},
  year={2008},
  institution={Los Alamos National Lab.(LANL), Los Alamos, NM (United States)}
}

@article{blackman2021xoshiro,
  title={Scrambled linear pseudorandom number generators},
  author={Blackman, David and Vigna, Sebastiano},
  journal={ACM Transactions on Mathematical Software (TOMS)},
  volume={47},
  number={4},
  pages={1--32},
  year={2021},
  publisher={ACM New York, NY}
}

@article{xorshift-rngs,
 title={Xorshift RNGs},
 volume={8},
 url={https://www.jstatsoft.org/index.php/jss/article/view/v008i14},
 doi={10.18637/jss.v008.i14},
 number={14},
 journal={Journal of Statistical Software},
 author={Marsaglia, George},
 year={2003},
 pages={1–6}
}

@misc{perf,
    title = {Perf: Linux profiling with performance counters},
    author = {Linux Community},
    howpublished = {\url{https://github.com/torvalds/linux/tree/master/tools/perf}},
    note = {Accessed: 2023-09-17}
}

@misc{intel-vtune-profiler,
    title = {Intel Vtune profiler},
    author = {Intel},
    howpublished = {\url{https://www.intel.com/content/www/us/en/developer/tools/oneapi/vtune-profiler.html}},
    note = {Accessed: 2023-09-17}
}

@misc{nvidia-nsight-systems,
    title = {NVIDIA Nsight Systems},
    author = {NVIDIA},
    howpublished = {\url{https://developer.nvidia.com/nsight-systems}},
    note = {Accessed: 2023-09-17}
}

@misc{nvidia-nsight-compute,
    title = {NVIDIA Nsight Compute},
    author = {NVIDIA},
    howpublished = {\url{https://developer.nvidia.com/nsight-compute}},
    note = {Accessed: 2023-09-17}
}

@inproceedings{gephi,
  title={Gephi: an open source software for exploring and manipulating networks},
  author={Bastian, Mathieu and Heymann, Sebastien and Jacomy, Mathieu},
  booktitle={Proceedings of the international AAAI conference on web and social media},
  volume={3},
  number={1},
  pages={361--362},
  year={2009}
}

@misc{curand-library,
    title = {NVIDIA cuRAND library},
    author = {NVIDIA},
    howpublished = {\url{https://developer.nvidia.com/curand}},
    note = {Accessed: 2023-09-17}
}

@inproceedings{yasin2014-topdown-approach,
  title={A top-down method for performance analysis and counters architecture},
  author={Yasin, Ahmad},
  booktitle={2014 IEEE International Symposium on Performance Analysis of Systems and Software (ISPASS)},
  pages={35--44},
  year={2014},
  organization={IEEE}
}

@inproceedings{pytorch,
 author = {Paszke, Adam and Gross, Sam and Massa, Francisco and Lerer, Adam and Bradbury, James and Chanan, Gregory and Killeen, Trevor and Lin, Zeming and Gimelshein, Natalia and Antiga, Luca and Desmaison, Alban and Kopf, Andreas and Yang, Edward and DeVito, Zachary and Raison, Martin and Tejani, Alykhan and Chilamkurthy, Sasank and Steiner, Benoit and Fang, Lu and Bai, Junjie and Chintala, Soumith},
 booktitle = {Advances in Neural Information Processing Systems},
 editor = {H. Wallach and H. Larochelle and A. Beygelzimer and F. d\textquotesingle Alch\'{e}-Buc and E. Fox and R. Garnett},
 pages = {},
 publisher = {Curran Associates, Inc.},
 title = {PyTorch: An Imperative Style, High-Performance Deep Learning Library},
 url = {https://proceedings.neurips.cc/paper_files/paper/2019/file/bdbca288fee7f92f2bfa9f7012727740-Paper.pdf},
 volume = {32},
 year = {2019}
}

@article{gibson2013survey,
  title={A survey of two-dimensional graph layout techniques for information visualisation},
  author={Gibson, Helen and Faith, Joe and Vickers, Paul},
  journal={Information visualization},
  volume={12},
  number={3-4},
  pages={324--357},
  year={2013},
  publisher={Sage Publications Sage UK: London, England}
}

@article{dwyer2009comparison,
  title={A comparison of user-generated and automatic graph layouts},
  author={Dwyer, Tim and Lee, Bongshin and Fisher, Danyel and Quinn, Kori Inkpen and Isenberg, Petra and Robertson, George and North, Chris},
  journal={IEEE transactions on visualization and computer graphics},
  volume={15},
  number={6},
  pages={961--968},
  year={2009},
  publisher={IEEE}
}

@inproceedings{blythe1995effect,
  title={The effect of graph layout on inference from social network data},
  author={Blythe, Jim and McGrath, Cathleen and Krackhardt, David},
  booktitle={International symposium on graph drawing},
  pages={40--51},
  year={1995},
  organization={Springer}
}

@inproceedings{purchase1997aesthetic,
  title={Which aesthetic has the greatest effect on human understanding?},
  author={Purchase, Helen},
  booktitle={International Symposium on Graph Drawing},
  pages={248--261},
  year={1997},
  organization={Springer}
}

@article{haleem2019evaluating,
  title={Evaluating the readability of force directed graph layouts: A deep learning approach},
  author={Haleem, Hammad and Wang, Yong and Puri, Abishek and Wadhwa, Sahil and Qu, Huamin},
  journal={IEEE computer graphics and applications},
  volume={39},
  number={4},
  pages={40--53},
  year={2019},
  publisher={IEEE}
}

@inproceedings{gansner2005graph-stress-majorization,
  title={Graph drawing by stress majorization},
  author={Gansner, Emden R and Koren, Yehuda and North, Stephen},
  booktitle={Graph Drawing: 12th International Symposium, GD 2004, New York, NY, USA, September 29-October 2, 2004, Revised Selected Papers 12},
  pages={239--250},
  year={2005},
  organization={Springer}
}

@article{kamada1989algorithm,
  title={An algorithm for drawing general undirected graphs},
  author={Kamada, Tomihisa and Kawai, Satoru and others},
  journal={Information processing letters},
  volume={31},
  number={1},
  pages={7--15},
  year={1989},
  publisher={Citeseer}
}

@article{le1986central,
  title={The central limit theorem around 1935},
  author={Le Cam, Lucien},
  journal={Statistical science},
  pages={78--91},
  year={1986},
  publisher={JSTOR}
}

@article{kwak2017central,
  title={Central limit theorem: the cornerstone of modern statistics},
  author={Kwak, Sang Gyu and Kim, Jong Hae},
  journal={Korean journal of anesthesiology},
  volume={70},
  number={2},
  pages={144},
  year={2017},
  publisher={Korean Society of Anesthesiologists}
}

@article{jacomy2014forceatlas2,
  title={ForceAtlas2, a continuous graph layout algorithm for handy network visualization designed for the Gephi software},
  author={Jacomy, Mathieu and Venturini, Tommaso and Heymann, Sebastien and Bastian, Mathieu},
  journal={PloS one},
  volume={9},
  number={6},
  pages={e98679},
  year={2014},
  publisher={Public Library of Science San Francisco, USA}
}

@article{fruchterman1991graph,
  title={Graph drawing by force-directed placement},
  author={Fruchterman, Thomas MJ and Reingold, Edward M},
  journal={Software: Practice and experience},
  volume={21},
  number={11},
  pages={1129--1164},
  year={1991},
  publisher={Wiley Online Library}
}

@article{hu2005efficient,
  title={Efficient, high-quality force-directed graph drawing},
  author={Hu, Yifan},
  journal={Mathematica journal},
  volume={10},
  number={1},
  pages={37--71},
  year={2005},
  publisher={Citeseer}
}

\includepdf[pages=-, pagecommand={\thispagestyle{plain}}]{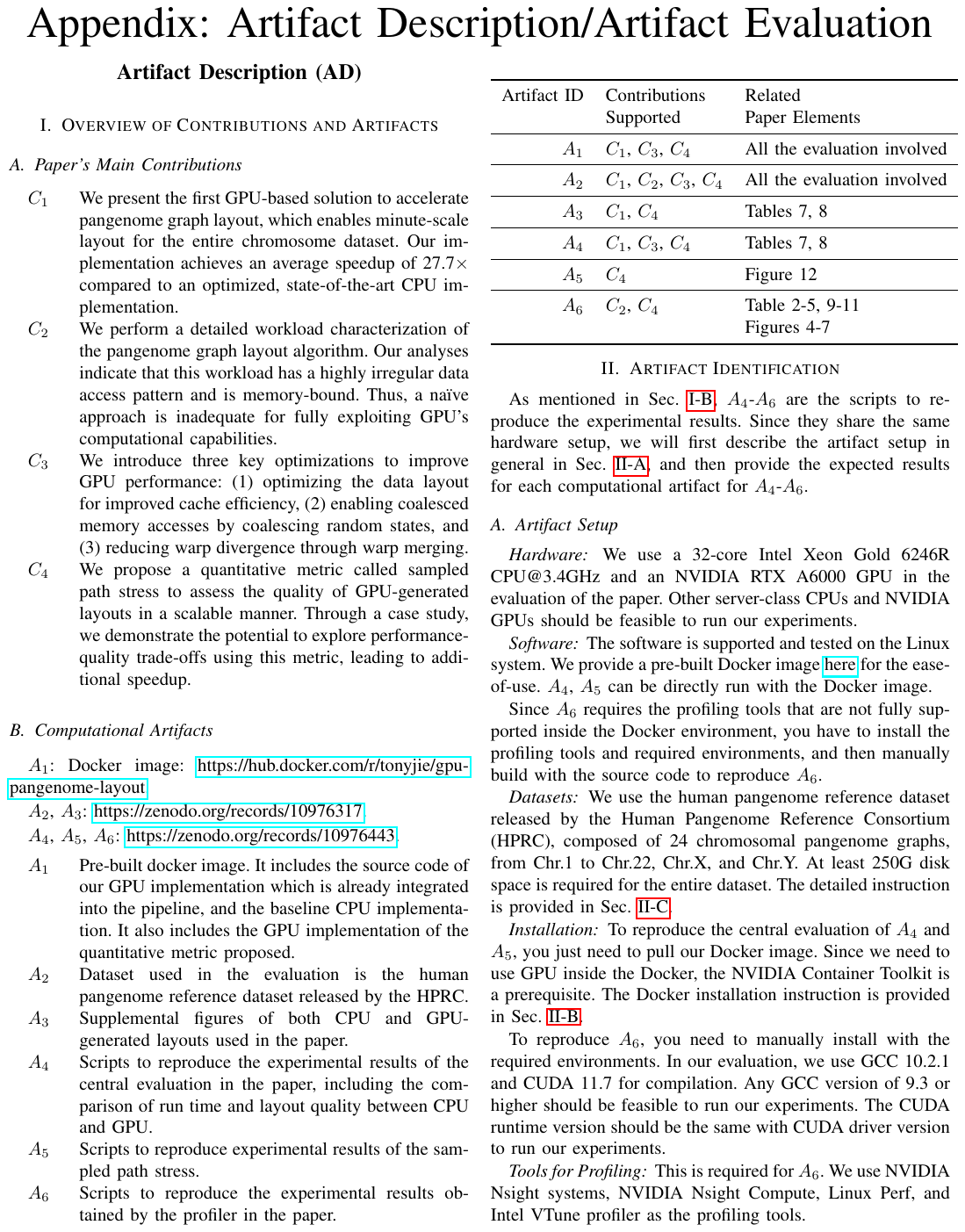}

\end{document}